\documentclass[prl,reprint,aps,letterpaper,floatfix,nobalancelastpage,twocolumn,superscriptaddress]{revtex4-1}
\pdfoutput=1
\usepackage{comment}
\nopagebreak
\usepackage{overpic}
\usepackage{rotating}
\usepackage{float}
\usepackage{amsmath}
\usepackage{amsfonts}
\usepackage{graphicx}
\usepackage{bm}
\usepackage{color}
\usepackage{braket}
\usepackage{tikz}
\usetikzlibrary{positioning}

\begin{document}

\title{Quantization of quasinormal modes for  open cavities and  plasmonic
cavity-QED }
\date{\today}

\author{Sebastian~Franke}
\email{sebastian.franke@tu-berlin.de}
	\affiliation{Technische Universit\"at Berlin, Institut f\"ur Theoretische Physik,
Nichtlineare Optik und Quantenelektronik, Hardenbergstra{\ss}e 36, 10623 Berlin, Germany}
\author{Stephen Hughes}
	\affiliation{Department of Physics, Engineering Physics and Astronomy, Queen's 		University, Kingston, Ontario K7L 3N6, Canada}
\author{Mohsen Kamandar Dezfouli}
	\affiliation{Department of Physics, Engineering Physics and Astronomy, Queen's University, Kingston, Ontario K7L 3N6, Canada}
\author{Philip Tr\o st Kristensen}
	\affiliation{Humboldt-Universit\"{a}t zu Berlin, AG Theoretische Optik \& Photonik, 12489 Berlin, Germany}
\author{ Kurt Busch}
	\affiliation{Humboldt-Universit\"{a}t zu Berlin, AG Theoretische Optik \& Photonik, 12489 Berlin, Germany}
	\affiliation{Max-Born-Institut, 12489 Berlin, Germany}
\author{Andreas Knorr}
	\affiliation{Technische Universit\"at Berlin, Institut f\"ur Theoretische Physik,
 Nichtlineare Optik und Quantenelektronik, Hardenbergstra{\ss}e 36, 10623 Berlin, Germany}
\author{Marten Richter}
	\affiliation{Technische Universit\"at Berlin, Institut f\"ur Theoretische Physik,
 Nichtlineare Optik und Quantenelektronik, Hardenbergstra{\ss}e 36, 10623 Berlin, Germany}

\begin{abstract}
We introduce a second quantization scheme 
based on quasinormal modes, which are the dissipative modes of leaky optical cavities and plasmonic resonators  
with complex eigenfrequencies.  
The theory  enables the construction of 
multi-plasmon/photon
Fock states for arbitrary  three-dimensional dissipative  
resonators 
and gives a solid understanding to the limits of 
phenomenological dissipative Jaynes-Cummings models.  
In the general case, we show 
how  different 
quasinormal modes 
interfere through an off-diagonal mode coupling and demonstrate how these results affect cavity-modified spontaneous emission. 
To illustrate the practical application of the theory, we show   
examples using  
a gold nanorod dimer and a hybrid dielectric-metal cavity structure.
\end{abstract}

\maketitle

Open cavity systems such as micropillars~\cite{micropillars,micropillars2,Reithmaier_Nature_432_197_2004},
photonic crystal cavities~\cite{Yoshie_Nature_432_200_2004,sh2007}, metal resonators~\cite{nanogold2,Akselrod2016,david,NanoMarten} and hybrid  metal-dielectric cavities~\cite{Barth2010,KamandarDezfouli2017} are of  interest for fundamental quantum optics and emerging technologies, including two-photon lasing~\cite{lasing}, spasing~\cite{spaser,PhysRevLett.Spaser}, and quantum information processing~\cite{quantinfo}. In such systems, one goal of quantum optics is to describe the  electric field as an operator  
associated with the creation and annihilation of photons or plasmons. One important example is the ``modes of the universe'' approach~\cite{motu1,motu2,motu3,motu4}; another is the 
Jaynes-Cummings (JC) model~\cite{Jaynes_ProcIEEE_51_89_1963, GirishBook1}, which describes multi-photon interactions between a quantized cavity mode and quantum emitters such as  molecules, two level atoms, or quantum dots~\cite{2levelsys1,gegg,RevModPhys.87.347}. A major restriction of the JC model is that the quantization procedure starts with the quantized modes of a closed cavity made from a lossless material. Such modes have real eigenfrequencies and can be normalized in a straight-forward way~\cite{GirishBook1}. For cavities with very low radiative loss, as usually quantified by a high quality factor $Q$, dissipation 
is often  
modelled through second order system-reservoir theory~\cite{GirishBook1} or quantum stochastical differential equations methods~\cite{QNoise,Gardiner1}. For low $Q$ cavities, however, or in the case of metallic resonators, such approaches are ambiguous, and  
the concept of a closed cavity is clearly problematic. 

While much progress has been made by treating the dissipation through the electromagnetic  
environment surrounding high $Q$ cavities as a bath, 
very little has been done in terms of quantized dissipative modes, especially in plasmonics. One  heuristic approach   has been to assume a phenomenological dissipative JC model, which for metals assumes parameters normally used for cavity modes, and where the total plasmon  
loss is treated phenomenologically~\cite{PhysRevA.82.043845}. Other notable approaches include the use of a 
projection operator applied to dielectric coupled-cavity systems~\cite{Dignam}, or treating
the electromagnetic  
environment by expansions 
in terms of pseudomodes~\cite{Hughes:18,hensenstrong} or  
quasinormal modes (QNMs)~\cite{PhysRevB.92.205420}. The QNMs 
offer tremendous insight and efficiency for describing electromagnetic scattering and semi-classical light-matter interaction~\cite{MDR1,muljarovPert,KristensenHughes,SauvanNorm,NormKristHughes,Lalanne_review}. Typically only a few QNMs and often just one QNM is needed, and it 
has recently been recognized that a quantum theory based on 
QNMs would represent a ``major milestone in quantum optics"~\cite{ACSAsger}. While some progress in this direction has been made for one-dimensional dielectric structures~\cite{2ndquanho,Severini}, these particular approaches do not  
lead to the construction of Fock states, which forms the 
natural basis for studying multi-plasmon/photon dynamics. 
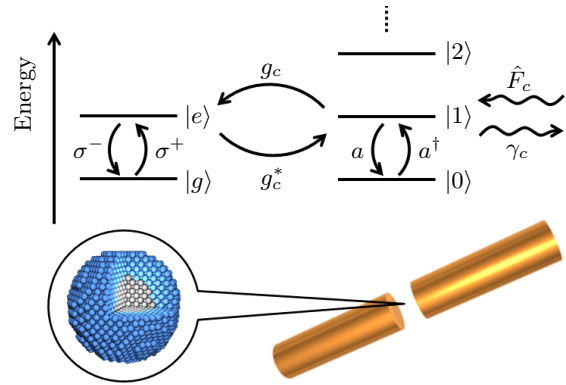
\begin{figure}[b]
 \vspace{-.5cm}
 \centering
	\begin{overpic}[width=0.90\columnwidth,angle=0]{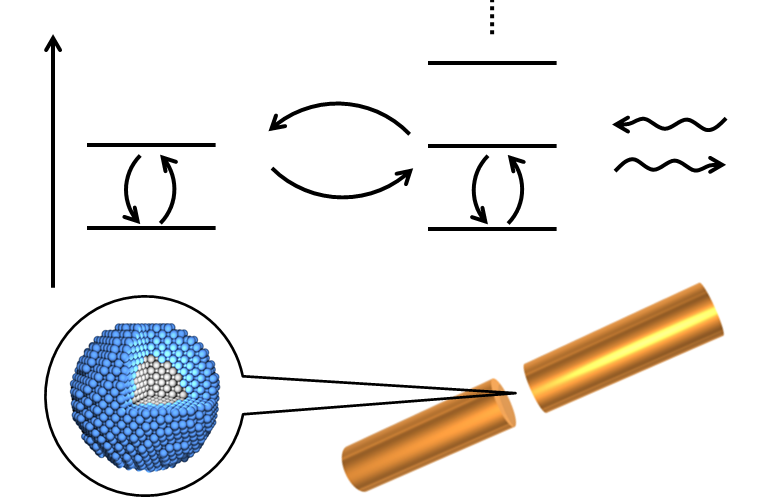}
	\put(0,43){\begin{sideways}Energy\end{sideways}}	
	\put(10,38.5){$\sigma^-$}	
	\put(24,38.5){$\sigma^+$}	
	\put(29,33.5){$|g\rangle$}	
	\put(29,44.5){$|e\rangle$}	
	\put(42,34){$g^*_c$}	
	\put(42,53){$g_c$}	
	\put(69,38.5){$a^{\dagger}$}	
	\put(57.5,38.5){$a$}	
	\put(73.5,33.5){$|0\rangle$}	
	\put(73.5,44.5){$|1\rangle$}	
	\put(73.5,55.5){$|2\rangle$}	
	\put(84,50.5){$\hat{F}_c$}	
	\put(84,39){$\gamma_c$}	
\end{overpic}
 \vspace{-0.2cm}
 \caption{Schematic of the QNM-JC model for a two level emitter, e.g., in the form of a colloidal quantum dot coupled to a single QNM of a plasmonic dimer of gold nanorods. 
The electronic states $|e\rangle$ and $|g\rangle$ are coupled to the plasmonic Fock states $|n\rangle$ via the coupling constant $g_c$. Quantum fluctuations associated with the electromagnetic dissipation 
 enter naturally through the operator $\hat{F}_c$.}
\label{fig0}
\vspace{-.1cm}
\end{figure}

In this Letter, we  present a rigorous quantization scheme for leaky optical cavities and plasmonic resonators using QNMs, which --- in contrast to typical approaches to lossy materials --- enables the construction of Fock states, as illustrated in Fig.~\ref{fig0}. We solve the main challenges related to the non-Hermitian nature of the
problem by introducing a symmetrization scheme for creation and annihilation operators associated with the resonator QNMs. Using the symmetrized operators, which satisfy canonical commutation relations, 
we derive the associated QNM-JC model for solving problems in multi-plasmon/photon quantum optics.
To illustrate the practical application of the theory, in the case of a single QNM, we first use an example of a gold nanorod resonator, as illustrated in Fig.~\ref{fig0}.  
Subsequently, we 
present the two-QNM-JC model and exemplify using a hybrid metal-dielectric cavity to highlight how  
modes quantum-mechanically interfere,  
leading to a dramatic breakdown of 
phenomenological dissipative  
JC models.

\textit{Theory}.---We consider the interaction between a two level emitter and the total electric field in the presence of a dispersive and absorptive, spatially inhomogeneous medium.    
The total Hamiltonian, $H_{\text{total}}{=}{H}_a {+} {H}_B{+}{H}_I$, based on the seminal approaches from \cite{HuttnerBarnett,A.Tip,Dung,Suttorp} is
\begin{align}
H_{\text{total}}=& \hbar \omega_a\sigma^+\sigma^- + \hbar\int {\rm d}\mathbf{r}\!\int_0^{\infty}\!{\rm d}\omega~\omega~\mathbf{b}^{\dagger}
(\mathbf{r},\omega)\cdot\mathbf{b}(\mathbf{r},\omega)\nonumber \\
&-\left[\sigma^+ \int_0^{\infty}{\rm d}\omega~\mathbf{d}_a\cdot\hat{\mathbf{E}}(\mathbf{r}_a,\omega)+\text{H.a.}\right],
\label{totham1}
\end{align}
where $\hbar$ is the reduced Planck constant, $\omega_a$ and $\mathbf{d}_a$ are the  resonance frequency and dipole moment of the emitter, respectively, $\sigma^\pm$ denote raising and lowering operators, and we use a dipole-field interaction in the rotating wave approximation;
the annihilation and creation operators $\mathbf{b}(\mathbf{r},\omega)$ and $\mathbf{b}^{\dagger}(\mathbf{r},\omega)$ act on joint excitations of the surrounding lossy media 
and electromagnetic degrees of freedom and 
satisfy  canonical commutation relations~\cite{Dung}. 
The  electric field operator 
fulfills the equation 
${\bm \nabla} {\times} {\bm \nabla} {\times}\hat{\mathbf{E}}(\mathbf{r},\omega){
-}k_0^2\epsilon(\mathbf{r},\omega)\hat{\mathbf{E}}(\mathbf{r},\omega)=
i\omega\mu_0 \hat {\bf j}_{\text{noise}}({\bf r},\omega)$, where $k_0=\omega/c$, $c=1/\sqrt{\mu_0\epsilon_0}$ is the speed of light, $\epsilon(\mathbf{r},\omega){=}\epsilon_R(\mathbf{r},\omega){+}i\epsilon_I(\mathbf{r},\omega)$ is a complex permittivity, describing passive media ($\epsilon_I(\mathbf{r},\omega){>}0$) and fulfilling the Kramers-Kronig relations, and $\hat{\mathbf{j}}_{\text{noise}}(\mathbf{r},\omega)=\omega
\sqrt{(\hbar\epsilon_0/\pi)\epsilon_{I}(\mathbf{r},\omega)}\mathbf{b}(\mathbf{r},\omega)$, with $\epsilon_0$ denoting the permittivity of free space~\cite{Dung,grunwel}. 
A formal solution  
is $\hat{\mathbf{E}}(\mathbf{r},\omega){=}i/(\omega\epsilon_0)\int {\rm d} \mathbf{r}' \mathcal{G}(\mathbf{r},\mathbf{r}',\omega)\hat{\mathbf{j}}_{\text{noise}}(\mathbf{r}',\omega),$ where $\mathcal{G}(\mathbf{r},\mathbf{r}',\omega)$ is the electric field  
Green's function, fulfilling $\nabla\times\nabla\times\mathcal{G}(\mathbf{r},\mathbf{r}',\omega)-k_0^2\epsilon(\mathbf{r},\omega)\mathcal{G}(\mathbf{r},\mathbf{r}',\omega)=k_0^2\delta(\mathbf{r}-\mathbf{r}')$ and a suitable radiation condition.
At this point, the quantization scheme from~\cite{Dung} provides already an intuitive picture for treating the system, but the frequency and spatial indices of $\mathbf{b}^{(\dagger)}(\mathbf{r},\omega)$ prevent numerical evaluations of the density matrix for applications beyond single photons (weak excitation). 

To derive creation and annihilation operators for photonic or plasmonic resonances, we begin with the vector-valued QNM eigenfunctions, $\tilde{\mathbf{f}}_{\mu}(\mathbf{r})$, defined from 
\begin{equation}
\nabla\times\nabla\times\tilde{\mathbf{f}}_{\mu}(\mathbf{r})-\frac{\tilde{\omega}_\mu^2}
{c^2}\epsilon(\mathbf{r},\tilde{\omega}_\mu)\tilde{\mathbf{f}}_{\mu}(\mathbf{r})=0,\label{HelmholtzQNM}
\end{equation}
with a suitable boundary condition, e.g., the Silver-M\"uller radiation condition~\cite{normaliz}.
The QNM eigenfrequencies $\tilde{\omega}_\mu{=}\omega_{\mu}{-}i\gamma_{\mu}$ are complex, with  $\gamma_{\mu}{>}0$ describing loss. For positions close to an electromagnetic resonator, the $\mathcal{G}(\mathbf{r},\mathbf{r}',\omega)$ can often be very accurately approximated by an expansion of only a few dominant QNMs of the form~\cite{MDR1,muljarovPert,KristensenHughes,Doost_PRA_87_043827_2013,SauvanNorm}
$\mathcal{G}_{\textit{\rm QNM}}(\mathbf{r},\mathbf{r}',\omega){=}\sum_{\mu} A_\mu(\omega) \tilde{\mathbf{f}}_{\mu}(\mathbf{r})\tilde{\mathbf{f}}_{\mu}(\mathbf{r}'),
$ where the QNMs are 
normalized~\cite{MDR1,muljarovPert,SauvanNorm,normaliz};  
for calculations in this work, we use the form $A_\mu(\omega){=}\omega/\left({2(\tilde{\omega}_{\mu}{-}\omega)}\right)$. 
Since the QNMs, $\tilde{\mathbf{f}}_{\mu}(\mathbf{r})$,  diverge in the far field,  we replace $\tilde{\mathbf{f}}_{\mu}(\mathbf{r})$ in $\mathcal{G}_{\textit{\rm QNM}}(\mathbf{r},\mathbf{r}',\omega)$ for positions outside the resonator geometry with a regularization based on a Dyson equation approach~\cite{Ge1}, $\tilde{\mathbf{F}}_{\mu}(\mathbf{r},\omega){=}\int_{V}{\rm d}\mathbf{r}'\mathcal{G}_{\rm B}(\mathbf{r},\mathbf{r}',\omega)\Delta\epsilon(\mathbf{r}',\omega)\tilde{\mathbf{f}}_{\mu}(\mathbf{r}')$, 
where $\mathcal{G}_{\rm B}$ is the   
homogeneous background medium Green's function, with constant permittivity $\epsilon_B$, $\Delta\epsilon(\mathbf{r}',\omega){=}\epsilon(\mathbf{r}',\omega){-}\epsilon_B$, and $V$ is the resonator geometry volume.
The electric field operator is $ \hat{\mathbf{E}}(\mathbf{r})\!=\!\int_0^\infty {\rm d}\omega\hat{\mathbf{E}}(\mathbf{r},\omega) {+}\text{H.a.}$ and can be expanded in basis functions of the Green's function as in Refs.~\cite{PlasQED,Rouss1,Rouss2,Rouss3}. Inspired by this elegant approach~\cite{PlasQED}, we use instead an expansion using few dominating (regularized) QNMs as in $\mathcal{G}_{\textit{\rm QNM}}(\mathbf{r},\mathbf{r}',\omega)$, so that the source field expression of the electric field operator can be rewritten,  $\hat{\mathbf{E}}(\mathbf{r}){=}i\sqrt{\hbar/(2\epsilon_0)}\sum_\mu\sqrt{\omega_\mu}\tilde{\mathbf{f}}_{\mu}(\mathbf{r}) \tilde{\alpha}_\mu+\text{H.a.}$,   for positions inside $V$, which allows us to define QNM operators $\tilde{\alpha}_\mu$=$\sqrt{2/(\pi\omega_\mu)}\int_0^{\infty}{\rm d}\omega A_\mu (\omega)\int {\rm d}\mathbf{r}\sqrt{\epsilon_I(\mathbf{r},\omega)}\,\tilde{\mathbf{f}}_{\mu}(\mathbf{r})\cdot\mathbf{b}(\mathbf{r},\omega)$; for positions outside $V$ in the spatial integral, 
$\tilde{\mathbf{f}}_{\mu}(\mathbf{r})$ is replaced by the regularized QNM $\tilde{\mathbf{F}}_{\mu}(\mathbf{r},\omega)$.
The QNM operators are formed by the integral over the oscillator operators $\mathbf{b}(\mathbf{r},\omega)$, that collectively form the QNM resonance. For most practical examples, the use of few (one or two) QNMs has been shown to provide very accurate results in the weak-coupling regime~\cite{Ge:14Dimer,KamandarDezfouli2017}; this is typically the case in cavity-QED systems using resonators.
Using the canonical commutation relations for $\mathbf{b}(\mathbf{r},\omega)$, the equal-time commutation relations for the QNM raising and lowering operators become $[\tilde{\alpha}_{\mu},
\tilde{\alpha}_{\eta}]\!=\![\tilde{\alpha}_{\mu}^{\dagger},\tilde{\alpha}_{\eta}^{\dagger}]\!=\!0$ and $[\tilde{\alpha}_{\mu},\tilde{\alpha}_{\eta}^{\dagger}]
{\equiv} S_{\mu\eta}$, in which 
\begin{equation}
S_{\mu\eta}{=}\int_0^{\infty}\!{\rm d}\omega \frac{2 A_\mu(\omega)A_\eta^*(\omega)}{\pi\sqrt{\omega_\mu\omega_\eta}}\left[S_{\mu\eta}^{\rm nrad}(\omega){+}S_{\mu\eta}^{\rm rad}(\omega)\right]\label{commutealpha},
\end{equation}
where  
$S_{\mu\eta}^{\rm nrad}(\omega)=\int_{V} {\rm d}\mathbf{r}\,\epsilon_I(\mathbf{r},\omega)\,
\tilde {\mathbf{f}}_{\mu}({\bf r})\cdot\tilde 
{\mathbf{f}}_{\eta}^*({\bf r})$ reflects absorption due to the resonator material, and 
$S_{\mu\eta}^{\rm rad}(\omega){=}c^2/(2i\omega^2)\int_{S_{V}}\! {\rm d}A_{\mathbf{s}} \{  [\hat{\mathbf{n}}_{\mathbf{s}}  {\times} (\nabla_{\mathbf{s}}
{\times}\tilde{\mathbf{F}}_\mu (\mathbf{s},\omega))] { \cdot }\tilde{\mathbf{F}}_\eta^*(\mathbf{s},\omega)
$
${-}[\hat{\mathbf{n}}_{\mathbf{s}} { \times}(\nabla_{\mathbf{s}}{\times}\tilde{\mathbf{F}}_\eta^* (\mathbf{s},\omega))]
{\cdot} \tilde{\mathbf{F}}_\mu(\mathbf{s},\omega)\}$ describes radiation leaving the system through the surface $S_V$ with the normal vector $\hat{\mathbf{n}}_{\mathbf{s}}$ pointing into $V$~\cite{SI}. The matrix  
$\left(\mathbf{S}\right)_{\mu\eta}$ is a Hermitian semi-positive definite overlap matrix between different QNMs $\mu,\eta$, and is  not a Kronecker delta as would be the case for closed, dielectric cavities. It is strictly positive definite if the modes are linearly independent, which we  assume here. Since non-canonical commutation relations prevent the construction of Fock states, we introduce new operators 
via a symmetrizing {\it orthonormalization} transformation~\cite{Chemie2}: 
\begin{equation}
a_{\mu}=\sum_{\nu}\big(\mathbf{S}^{-\frac{1}{2}}\big)_{\mu\nu}\tilde{\alpha}_{\nu},~
a^{\dagger}_\mu=
\sum_{\nu}\big(\mathbf{S}^{-\frac{1}{2}}\big)_{\nu\mu}\tilde{\alpha}_{\nu}^{\dagger},
\label{newalphadagger}
\end{equation}
yielding 
$[a_{\mu},a_{\eta}^{\dagger}]{=}\delta_{\mu\eta}$. The operators    
$a_\mu$ and $a_\mu^{\dagger}$ are therefore proper 
annihilation and creation operators for obtaining  
Fock states  $|\mathbf{n}\rangle\!\equiv\!|n_1,n_2,...\rangle$ from  
the
 vacuum state $|0\rangle$. The electric field expressed by $a_{\mu},a_{\mu}^{\dagger}$ is then
\begin{equation}
\hat{\mathbf{E}}(\mathbf{r})=i\sqrt{\frac{\hbar }{2\epsilon_0  }}\sum_{\mu}\sqrt{\omega_\mu}\, \tilde{\mathbf{f}}_{\mu}^{\rm s}(\mathbf{r})  a_\mu + \text{H.a.},\label{Esymm}
\end{equation}
in the desired basis and with {\it symmetrized} QNM functions $\tilde{\mathbf{f}}^{\rm s}_{\mu}(\mathbf{r})$=$\sum_{\nu}(\mathbf{S}^{\frac{1}{2}})_{\nu\mu} \sqrt{\omega_\nu /\omega_\mu}\,\tilde{\mathbf{f}}_{\nu}(\mathbf{r})$. For a single QNM, $\mu\!=\!\nu\!=\!c$, we get  $\hat{\mathbf{E}}(\mathbf{r})$=$i\sqrt{\hbar\omega_c/(2\epsilon_0)}
\tilde{\mathbf{f}}_{c}^{\rm s}(\mathbf{r})  a + \text{H.a.,}$ with $\tilde{\mathbf{f}}_{c}^{\rm s}(\mathbf{r}){=}\sqrt{S_c}
\,\tilde{\mathbf{f}}_{c}(\mathbf{r})$, $S_{c}=S_{cc}$ and $a\equiv a_c$. 
\begin{figure}[t]
 \centering
 \includegraphics[width=0.99\columnwidth,angle=0]{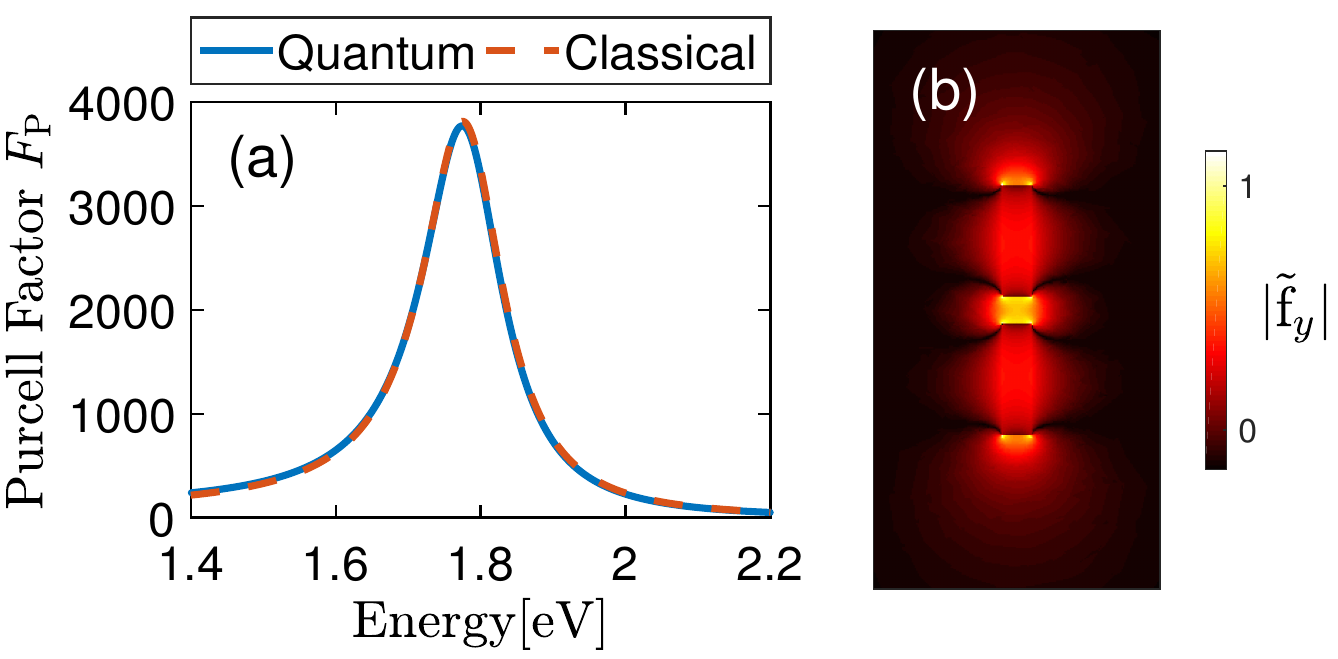}
 \vspace{-0.2cm}
 \caption{{
 (a) Purcell factor $F_\text{P}$ as a function of frequency for the plasmonic dimer in Fig.~\ref{fig0}. Solid and dashed curves show the results of the QNM-JC model and a semi-classical approach, using a single QNM Green's function approximation, respectively. (b) Normalized spatial profile of the QNM of interest with $\tilde \omega_c({\rm eV}) {=} 1.7786  {-} 0.0677i$, corresponding to $Q{=}\omega_c/(2\gamma_c) \approx 13$}.}
\label{fig1}
\end{figure}  
The dynamics of the operators $a_{\mu}$ are governed by the Heisenberg equations of motion~\cite{SI},
\begin{equation}
\frac{\text{d}}{\text{d}t}a_{\mu}\!=\!-\frac{i}{\hbar}[a_\mu,H_{\text{sys}}]-\sum_\eta \chi_{\mu\eta}^{(-)}a_{\eta}+\hat{F}_\mu,\label{QLE}
\end{equation}
where $H_{\text{sys}}{=}H_{\text{em}}{+}H_a{+}H_{I}$  is the effective system Hamiltonian in the symmetrized  basis, and
$H_{\text{em}}{=}\hbar\sum_{\mu\eta}\chi_{\mu\eta}^{(+)}a^{\dagger}_{\mu}a_{\eta}$ is the electromagnetic  
part with non-diagonal terms $\eta{\neq}\mu$, that includes coupling between different symmetrized QNMs $\mu,\eta$~\cite{2ndquanho}; $H_I{=}{-}i\hbar \sum_{\mu}g_{\mu}\sigma^+a_{\mu}{+}\text{H.a.}$ is derived from   $H_I$   in Eq.~\eqref{totham1} by inserting Eq.~\eqref{Esymm};
this yields an emitter-QNM dipole-field interaction in the symmetrized basis  $g_{\mu}$=$\sum_\eta\big(\mathbf{S}^{\frac{1}{2}})_{\eta\mu}\tilde{g}_\eta $, with $\tilde{g}_{\mu}$=$\sqrt{\omega_\mu/(2\epsilon_0 \hbar)}\, \mathbf{d}_a {\cdot} \tilde{\mathbf{f}}_\mu (\mathbf{r}_a)$.  In contrast to a phenomenological approach, complex QNM eigenfrequencies and complex eigenfunctions are used to obtain $g_{\mu}$.
The  coupling between different QNMs is induced by the symmetrizing transformation, Eq.~\eqref{newalphadagger}, and the coupling constant $\chi_{\mu\eta}^{(+)}$ is given via $\chi_{\mu\eta}^{(+)}{=}\frac{1}{2}
(\chi_{\mu\eta}{+}\chi_{\eta\mu}^*)$, where ${\chi}_{\mu\eta}{=}\sum_{\nu}
(\mathbf{S}^{-\frac{1}{2}})_{\mu\nu}\tilde{\omega}_\nu\big(\mathbf{S}^{\frac{1}{2}})_{\nu\eta}$. The two additional terms in Eq.~\eqref{QLE} account for dissipation of energy 
through $\chi_{\mu\eta}^{(-)}{=}\frac{i}{2}(\chi_{\mu\eta}$-$\chi_{\eta\mu}^*)$ and coupling to a noise term $\hat{F}_{\mu}=\sum_{\nu}\big(\mathbf{S}^{-\frac{1}{2}}\big)_{\mu\nu}\int_0^{\infty} {\rm d}\omega\,C_\nu(\omega) \int {\rm d}\mathbf{r}\sqrt{\omega\epsilon_I(\mathbf{r},\omega)}
\,\tilde{\mathbf{f}}_{\nu}(\mathbf{r})\cdot\mathbf{b}(\mathbf{r},\omega)$, with $C_\nu(\omega){=}i\sqrt{\omega/(2\pi\omega_\nu)}$. The presence of the noise term preserves the commutation relation $[a_{\mu},a^{\dagger}_\eta]{=}\delta_{\mu\eta}$ temporally by exactly counteracting the damping due to dissipation.
Equation~\eqref{QLE} has the form of a Langevin equation with noise operators $\hat{F}_{\mu}$~\cite{QNoise,Gardiner1}. In the following, we use it to set up the QNM-JC model and derive the associated quantum master equations for the illustrative cases of one and two QNMs; we refer to Ref.~\onlinecite{SI} for the general case. 

(I) \textit{One-QNM-JC model}.\ 
We consider the material system from Fig.~\ref{fig0} consisting of an emitter at the position $\mathbf{r}_a$, directly in the center of a plasmonic dimer of nanorods supporting a single QNM with index $c$, in which case
$H_{\text{em}}$ takes the 
simplified form  
$H_{\text{em}}{=}\hbar\omega_c a^{\dagger}a$, and the  
interaction Hamiltonian becomes 
$H_I{=}{-}i\hbar(g_c\sigma^+a{-}\text{H.a.})$, with $g_c{=}\sqrt{S_{c}}\,\tilde{g}_c$ and $\chi_{\mu\eta}^{(-)}{=} 
{-}\text{Im}(\tilde{\omega}_c){=}\gamma_c$.
All system operators evolve according to a quantum Langevin equation similar to Eq.~\eqref{QLE}. Employing 
stochastic Ito/Stratonovich calculus~\cite{Gardiner1}, we can bring these equations into
a Lindblad master equation form. To this end, we treat the quantum noise operator $\hat{F}_c$ as an input field, which represents 
white noise of a reservoir with temperature $T$=$0\,{\rm K}$~\cite{SI}. After some algebra, we find the one-QNM master equation
\begin{equation}
\partial_t \rho = -\frac{i}{\hbar}\left[H_{\rm sys},\rho\right]+
\gamma_c\left(2a\rho a^\dagger-a^\dagger a\rho-\rho a^\dagger a\right).\label{masterQNMV2}
\end{equation}
For a more detailed interpretation of the normalization $S_c$, it is instructive to consider the cavity-modified spontaneous emission  rate in  the bad cavity limit. We adiabatically eliminate the electromagnetic degrees of freedom 
from  Eq.~\eqref{masterQNMV2}, to obtain a master equation for the emitter density matrix alone, consisting of the dissipator term $\mathcal{L}[\sigma^-]\rho$=$\Gamma
 \left(2\sigma^-\rho\sigma^+ -\sigma^+\sigma^-\rho -\rho\sigma^+\sigma^-\right)$, with  
spontaneous emission rate
$\Gamma 
=\gamma_c|g_c|^2/(\Delta_{ca}^2 + \gamma_c^2) $  and 
detuning $\Delta_{ca}$=$\omega_c\!-\!\omega_a$~\cite{SI}. 
In cases where the single QNM expansion is a good approximation to the Green's function throughout the entire resonator volume, we expect to recover the semi-classical result
$\Gamma 
=(2/\hbar\epsilon_0)\mathbf{d}_a\cdot\text{Im}\left(\mathcal{G}(\mathbf{r}_a,\mathbf{r}_a,\omega)\right)\cdot\mathbf{d}_a$, i.e., the local density of states modelled via the photonic Green's function in a QNM approximation~\cite{NormKristHughes,Ge:14Dimer}. 
In the present case of the plasmonic dimer, we 
find a very good agreement, as 
seen in Fig.~\ref{fig1}, showing 
the Purcell factor $F_{\rm P}$=$\Gamma 
/\Gamma^0 
$, where $\Gamma^0 
$
is the spontaneous emission rate in a homogeneous medium.
Although the agreement in Fig.~\ref{fig1} is already striking (especially given the completely different nature of the calculations~\cite{SI}), we remark that the restriction to few dominant QNMs in the QNM-JC model, when applied to spontaneous emission, are generally different, and typically less accurate, than the use of the same approximation to the Green's function in a semi-classical approach. Whereas the latter relies only on the expansion at a single point, the QNM-JC model is based on integrals of the QNMs throughout the resonator material to obtain $S_{c}$.  
For the plasmonic dimer, we find 
$S_{c}^{\rm nrad} {=} 0.58, S_{c}^{\rm rad} {=} 0.40$. 
In addition, $S_{c}^{\rm nrad}$ and $S_{c}^{\rm rad}$ yield  
the non-radiative and radiative  
beta factor via $\beta^{\rm nrad}=S_{c}^{\rm nrad}/S_{c}$ and $\beta^{\rm rad}=S_{c}^{\rm rad}/S_{c}$. 
See~\cite{SI} for details of the QNM calculation, $\tilde {\mathbf{f}}_\mu({\bf r}_a)$ and material parameters.

\begin{figure}[t]
 \centering
 \includegraphics[width=0.95\columnwidth,angle=0]{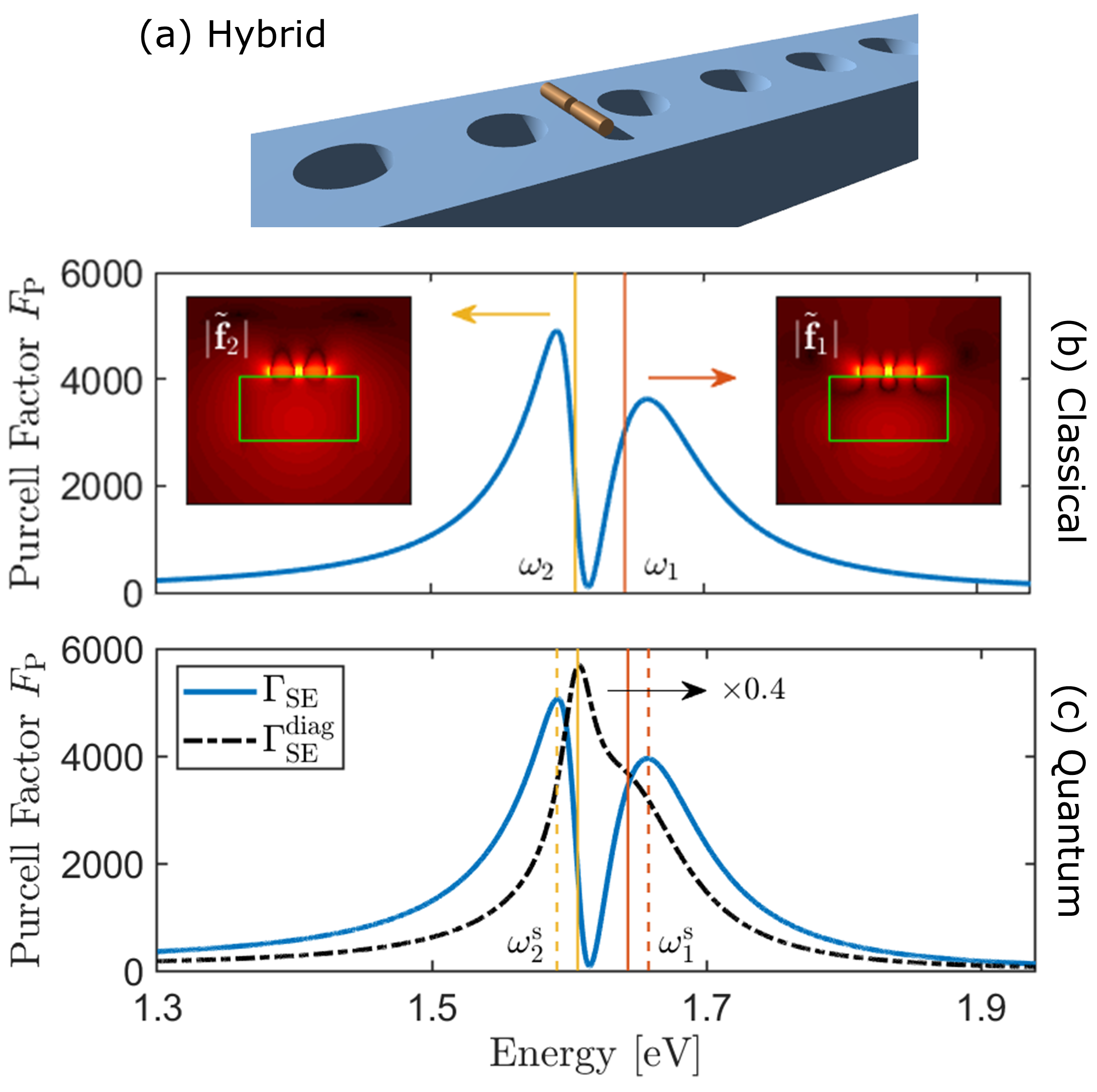}
 \vspace{-0.2cm}
 \caption{{(a) Gold  dimer on top of a photonic crystal cavity, supporting two overlapping QNMs with frequencies   $\tilde \omega_2({\rm eV}) {=} 1.6063 {-} 0.0145i$ and $\tilde \omega_1({\rm eV}) {=}  1.6428 {-} 0.0548i$ (mode 1 originates from the dimer).   (b) QNM profiles and semi-classical Purcell factor 
as a function of energy. (c) QNM-JC Purcell factor 
with diagonal contributions $\Gamma 
^{\rm diag}$ (black dashed, scaled) and the full emission rate $\Gamma
=\Gamma 
^{\rm diag}+\Gamma 
^{\rm ndiag}$ (solid blue). 
Vertical solid and dashed lines, show the shifted and original eigenfrequencies, 
respectively.}}
\label{fig2}
\end{figure}

(II) \textit{Two-QNM-JC model.} 
We next discuss a case where  cross-terms $\chi_{\mu\eta}$ of  two QNMs $\mu,\eta{=}1,2$ cause interference effects, clearly not available in phenomenological quantization approaches. Starting again from the quantum Langevin equation in Eq.~\eqref{QLE}, we derive a Lindblad master equation analogue to the one mode case, using the additional assumptions~\cite{Gardiner1} that the two input fields associated with $\hat{F}_\mu$ ($\mu=1,2$) are independent from each other 
and that the real parts of the eigenfrequencies $\omega_1,\omega_2$ are not degenerate~\cite{Lax}. Again, following the approach of Ref.~\onlinecite{Gardiner1}, we now obtain the two-QNM master equation
\begin{equation}
\partial_t \rho=
-\frac{i}{\hbar}\left[H_{\text{sys}},\rho\right]+\mathcal{L}[\textbf{a}]\rho\label{masterQNM2modes},
\end{equation}
where $\omega_\mu$ are no longer eigenvalues of 
the electromagnetic part of the 
Hamiltonian, since an inter-mode coupling appears. Instead, a pair of shifted eigenfrequencies $\omega_\mu^{\rm s}$ is formed 
(see Fig.~\ref{fig2}(c)). 
We stress that the Lindblad dissipator
$\mathcal{L}[\textbf{a}]\rho{=}\sum_{\mu,\eta}\chi_{\mu\eta}^{(-)}(2a_\eta\rho a^\dagger_\mu{-}
a^\dagger_\mu a_\eta\rho {-}\rho a_\mu^\dagger a_\eta)$ 
contains also
processes with interacting QNMs $\mu{\neq}\eta$.

Although the above off-diagonal coupling may seem unusual, 
it is known that a significant mode interference, such as a ``Fano-type" resonance, can occur because of the different phase terms of  overlapping QNMs~\cite{deLasson,KamandarDezfouli2017}. In the
QNM-JC model, this interference is captured by the off-diagonal terms, as illustrated in Fig.~\ref{fig2}, where we study 
the electromagnetic response of the metal dimer from (I) on top of a high-Q photonic crystal cavity (see Fig.~\ref{fig2}(a)).  
Figure~\ref{fig2}(b)  
shows the two QNMs of interest and the semi-classical result of the Purcell factor as calculated using a two-QNM approximation \cite{KamandarDezfouli2017,SI};  
Fig.~\ref{fig2}(c) shows  
the corresponding results of the QNM-JC model in this pronounced QNM coupling regime
\footnote{The calculated $S$ factors are: $S_{11}{=}0.7485 + 0.7356$, $S_{22}{=}0.8006+0.4409$, $S_{12}{ = }({-}0.0043 {-} 0.5574i)+(0.0579{-} 0.4137i)$, and $S_{21}{=}S^*_{12}$, where first and second parts denote $S_{\mu\eta}^{\rm nrad}$ and $S_{\mu\eta}^{\rm rad}$, respectively.}. The 
system parameters indicate the bad cavity limit, where 
the QNM-JC master equation consists of a Lindblad dissipator for spontaneous emission of the form 
$\mathcal{L}[\sigma^-]\rho{=}\Gamma 
\left(2\sigma^-\rho\sigma^+ {-}\sigma^+\sigma^-\rho {-}\rho\sigma^+\sigma^-\right)$,
in which 
$\Gamma
{=}\Gamma^{\rm diag} 
{+}\Gamma^{\rm ndiag} 
$ with a diagonal contribution $\Gamma^{\rm diag} 
= \sum_{\mu} S_{\mu\mu}|\tilde{g}_\mu|^2\gamma_\mu/(\Delta_{\mu a}^2 {+} \gamma_\mu^2)$ and a non-diagonal contribution $\Gamma^{\rm ndiag}
{=}\sum_{\mu,\eta\neq\mu}\tilde{g}_\mu S_{\mu\eta}\tilde{g}_{\eta}^{*}K_{\mu\eta}$, which is here expressed in terms of  
the coupling matrix $K_{\mu\eta}$=$(i(\omega_\mu{-}\omega_{\eta}){+}\gamma_\mu{+}\gamma_{\eta})/(2(\Delta_{\mu a}{-}i \gamma_\mu)(\Delta_{\eta a}{+}i\gamma_{\eta}))$ \cite{SI}. Comparing the results 
in Fig.~\ref{fig2}, one sees that the two-QNM-JC model 
recovers the result of the semi-classical calculation, 
including the pronounced Fano-type interference effect. In  
a phenomenological dissipative two-mode JC model, the Lindblad dissipator  is simply $\mathcal{L}\left[\sigma^-\right]\rho=\sum_{i}\Gamma 
^i\left(2\sigma^-\rho\sigma^+ {-}\sigma^+\sigma^-\rho {-}\rho\sigma^+\sigma^-\right)$ and $\Gamma
^i {=} |g_i|^2\gamma_i/(\Delta_{ia}^2 {+} \gamma_i^2)$ is the diagonal cavity-modified  rate for each of the two modes. Clearly,  such a model cannot produce  
the aforementioned interference effect, as illustrated by the black dashed curve in Fig.~\ref{fig2}(c).

By construction, the symmetrized raising and lowering operators fulfill all requirements for use in the construction of a Fock space. Combined with the fact that the one- and two-QNM-JC master equations recover the semi-classical results in the single excitation subspace, where a direct comparison to reference calculations is possible, we consider the approach rigorous enough that one can apply the QNM-JC model also to problems in multi-plasmon/photon Fock spaces. Although we have connected to the bad cavity limit, the QNM master equation now allows one to explore multi-photon dynamics, which will be the subject of future work.

In conclusion, by use of 
a symmetrization procedure, we have introduced creation and annihilation operators, allowing the construction of QNM Fock states and the derivation of a physically meaningful and intuitive  QNM-JC model for use in dissipative cavity-QED valid for arbitrary dissipative structures. We have shown  example applications of the theory for a plasmonic  
dimer to verify that the QNM-JC model recovers the semi-classical result in Purcell factor calculations. Finally, we discussed the highly non-trivial case of a two-QNM-JC model, where interference effects cannot be neglected. In this case, the model recovers the semi-classical result only because of off-diagonal coupling terms, which are not present in phenomenological dissipative  
JC models. Contrary to phenomenological approaches, all parameters entering the model are rigorously defined and can be calculated by use of the relevant QNMs. The model thus provides a solid foundation for the use of the JC model and a rigorous extension of the model to several modes and possibly dissipative materials, in which the quantized ``cavity modes" of optical cavities or plasmonic resonators appear as  
linear combinations of the associated QNMs.

We acknowledge funding from 
the Deutsche Forschungsgemeinschaft through SFB 951 HIOS (B12/B10) and project BR1528/8-2 for support, and from the Natural Sciences and Engineering Research Council of Canada. S.F. acknowledges support from the School of Nanophotonics (SFB 787). This project has also received funding from the
European Unions Horizon 2020 research and innovation
programme under Grant Agreement No. 734690 (SONAR).


\begin{thebibliography}{58}%
\makeatletter
\providecommand \@ifxundefined [1]{%
 \@ifx{#1\undefined}
}%
\providecommand \@ifnum [1]{%
 \ifnum #1\expandafter \@firstoftwo
 \else \expandafter \@secondoftwo
 \fi
}%
\providecommand \@ifx [1]{%
 \ifx #1\expandafter \@firstoftwo
 \else \expandafter \@secondoftwo
 \fi
}%
\providecommand \natexlab [1]{#1}%
\providecommand \enquote  [1]{``#1''}%
\providecommand \bibnamefont  [1]{#1}%
\providecommand \bibfnamefont [1]{#1}%
\providecommand \citenamefont [1]{#1}%
\providecommand \href@noop [0]{\@secondoftwo}%
\providecommand \href [0]{\begingroup \@sanitize@url \@href}%
\providecommand \@href[1]{\@@startlink{#1}\@@href}%
\providecommand \@@href[1]{\endgroup#1\@@endlink}%
\providecommand \@sanitize@url [0]{\catcode `\\12\catcode `\$12\catcode
  `\&12\catcode `\#12\catcode `\^12\catcode `\_12\catcode `\%12\relax}%
\providecommand \@@startlink[1]{}%
\providecommand \@@endlink[0]{}%
\providecommand \url  [0]{\begingroup\@sanitize@url \@url }%
\providecommand \@url [1]{\endgroup\@href {#1}{\urlprefix }}%
\providecommand \urlprefix  [0]{URL }%
\providecommand \Eprint [0]{\href }%
\providecommand \doibase [0]{http://dx.doi.org/}%
\providecommand \selectlanguage [0]{\@gobble}%
\providecommand \bibinfo  [0]{\@secondoftwo}%
\providecommand \bibfield  [0]{\@secondoftwo}%
\providecommand \translation [1]{[#1]}%
\providecommand \BibitemOpen [0]{}%
\providecommand \bibitemStop [0]{}%
\providecommand \bibitemNoStop [0]{.\EOS\space}%
\providecommand \EOS [0]{\spacefactor3000\relax}%
\providecommand \BibitemShut  [1]{\csname bibitem#1\endcsname}%
\let\auto@bib@innerbib\@empty
%</preamble>
\bibitem [{\citenamefont {Reitzenstein}\ \emph {et~al.}(2007)\citenamefont
  {Reitzenstein}, \citenamefont {Hofmann}, \citenamefont {Gorbunov},
  \citenamefont {Strau\ss}, \citenamefont {Kwon}, \citenamefont {Schneider},
  \citenamefont {L\"offler}, \citenamefont {H\"ofling}, \citenamefont {Kamp},\
  and\ \citenamefont {Forchel}}]{micropillars}%
  \BibitemOpen
  \bibfield  {author} {\bibinfo {author} {\bibfnamefont {S.}~\bibnamefont
  {Reitzenstein}}, \bibinfo {author} {\bibfnamefont {C.}~\bibnamefont
  {Hofmann}}, \bibinfo {author} {\bibfnamefont {A.}~\bibnamefont {Gorbunov}},
  \bibinfo {author} {\bibfnamefont {M.}~\bibnamefont {Strau\ss}}, \bibinfo
  {author} {\bibfnamefont {S.~H.}\ \bibnamefont {Kwon}}, \bibinfo {author}
  {\bibfnamefont {C.}~\bibnamefont {Schneider}}, \bibinfo {author}
  {\bibfnamefont {A.}~\bibnamefont {L\"offler}}, \bibinfo {author}
  {\bibfnamefont {S.}~\bibnamefont {H\"ofling}}, \bibinfo {author}
  {\bibfnamefont {M.}~\bibnamefont {Kamp}}, \ and\ \bibinfo {author}
  {\bibfnamefont {A.}~\bibnamefont {Forchel}},\ }\href@noop {} {\bibfield
  {journal} {\bibinfo  {journal} {Appl. Phys. Lett.}\ }\textbf {\bibinfo
  {volume} {90}},\ \bibinfo {pages} {251109} (\bibinfo {year}
  {2007})}\BibitemShut {NoStop}%
\bibitem [{\citenamefont {Bajoni}\ \emph {et~al.}(2008)\citenamefont {Bajoni},
  \citenamefont {Senellart}, \citenamefont {Wertz}, \citenamefont {Sagnes},
  \citenamefont {Miard}, \citenamefont {Lema\^{\i}tre},\ and\ \citenamefont
  {Bloch}}]{micropillars2}%
  \BibitemOpen
  \bibfield  {author} {\bibinfo {author} {\bibfnamefont {D.}~\bibnamefont
  {Bajoni}}, \bibinfo {author} {\bibfnamefont {P.}~\bibnamefont {Senellart}},
  \bibinfo {author} {\bibfnamefont {E.}~\bibnamefont {Wertz}}, \bibinfo
  {author} {\bibfnamefont {I.}~\bibnamefont {Sagnes}}, \bibinfo {author}
  {\bibfnamefont {A.}~\bibnamefont {Miard}}, \bibinfo {author} {\bibfnamefont
  {A.}~\bibnamefont {Lema\^{\i}tre}}, \ and\ \bibinfo {author} {\bibfnamefont
  {J.}~\bibnamefont {Bloch}},\ }\href {\doibase 10.1103/PhysRevLett.100.047401}
  {\bibfield  {journal} {\bibinfo  {journal} {Phys. Rev. Lett.}\ }\textbf
  {\bibinfo {volume} {100}},\ \bibinfo {pages} {047401} (\bibinfo {year}
  {2008})}\BibitemShut {NoStop}%
\bibitem [{\citenamefont {Reithmaier}\ \emph {et~al.}(2004)\citenamefont
  {Reithmaier}, \citenamefont {Sek}, \citenamefont {L\"offler}, \citenamefont
  {Hofmann}, \citenamefont {Kuhn}, \citenamefont {Reitzenstein}, \citenamefont
  {Keldysh}, \citenamefont {Kulakovskii}, \citenamefont {Reinecke},\ and\
  \citenamefont {Forchel}}]{Reithmaier_Nature_432_197_2004}%
  \BibitemOpen
  \bibfield  {author} {\bibinfo {author} {\bibfnamefont {J.~P.}\ \bibnamefont
  {Reithmaier}}, \bibinfo {author} {\bibfnamefont {G.}~\bibnamefont {Sek}},
  \bibinfo {author} {\bibfnamefont {A.}~\bibnamefont {L\"offler}}, \bibinfo
  {author} {\bibfnamefont {C.}~\bibnamefont {Hofmann}}, \bibinfo {author}
  {\bibfnamefont {S.}~\bibnamefont {Kuhn}}, \bibinfo {author} {\bibfnamefont
  {S.}~\bibnamefont {Reitzenstein}}, \bibinfo {author} {\bibfnamefont {L.~V.}\
  \bibnamefont {Keldysh}}, \bibinfo {author} {\bibfnamefont {V.~D.}\
  \bibnamefont {Kulakovskii}}, \bibinfo {author} {\bibfnamefont {T.~L.}\
  \bibnamefont {Reinecke}}, \ and\ \bibinfo {author} {\bibfnamefont
  {A.}~\bibnamefont {Forchel}},\ }\href@noop {} {\bibfield  {journal} {\bibinfo
   {journal} {Nature}\ }\textbf {\bibinfo {volume} {432}},\ \bibinfo {pages}
  {197} (\bibinfo {year} {2004})}\BibitemShut {NoStop}%
\bibitem [{\citenamefont {Yoshie}\ \emph {et~al.}(2004)\citenamefont {Yoshie},
  \citenamefont {Scherer}, \citenamefont {Hendrickson}, \citenamefont
  {Khitrova}, \citenamefont {Gibbs}, \citenamefont {Rupper}, \citenamefont
  {Ell}, \citenamefont {Shchekin},\ and\ \citenamefont
  {Deppe}}]{Yoshie_Nature_432_200_2004}%
  \BibitemOpen
  \bibfield  {author} {\bibinfo {author} {\bibfnamefont {T.}~\bibnamefont
  {Yoshie}}, \bibinfo {author} {\bibfnamefont {A.}~\bibnamefont {Scherer}},
  \bibinfo {author} {\bibfnamefont {J.}~\bibnamefont {Hendrickson}}, \bibinfo
  {author} {\bibfnamefont {G.}~\bibnamefont {Khitrova}}, \bibinfo {author}
  {\bibfnamefont {H.~M.}\ \bibnamefont {Gibbs}}, \bibinfo {author}
  {\bibfnamefont {G.}~\bibnamefont {Rupper}}, \bibinfo {author} {\bibfnamefont
  {C.}~\bibnamefont {Ell}}, \bibinfo {author} {\bibfnamefont {O.~B.}\
  \bibnamefont {Shchekin}}, \ and\ \bibinfo {author} {\bibfnamefont {D.~G.}\
  \bibnamefont {Deppe}},\ }\href@noop {} {\bibfield  {journal} {\bibinfo
  {journal} {Nature}\ }\textbf {\bibinfo {volume} {432}},\ \bibinfo {pages}
  {200} (\bibinfo {year} {2004})}\BibitemShut {NoStop}%
\bibitem [{\citenamefont {Manga~Rao}\ and\ \citenamefont
  {Hughes}(2007)}]{sh2007}%
  \BibitemOpen
  \bibfield  {author} {\bibinfo {author} {\bibfnamefont {V.~S.~C.}\
  \bibnamefont {Manga~Rao}}\ and\ \bibinfo {author} {\bibfnamefont
  {S.}~\bibnamefont {Hughes}},\ }\href {\doibase 10.1103/PhysRevLett.99.193901}
  {\bibfield  {journal} {\bibinfo  {journal} {Phys. Rev. Lett.}\ }\textbf
  {\bibinfo {volume} {99}},\ \bibinfo {pages} {193901} (\bibinfo {year}
  {2007})}\BibitemShut {NoStop}%
\bibitem [{\citenamefont {{Kulakovich}}\ \emph {et~al.}(2002)\citenamefont
  {{Kulakovich}}, \citenamefont {{Strekal}}, \citenamefont {{Yaroshevich}},
  \citenamefont {{Maskevich}}, \citenamefont {{Gaponenko}}, \citenamefont
  {{Nabiev}}, \citenamefont {{Woggon}},\ and\ \citenamefont
  {{Artemyev}}}]{nanogold2}%
  \BibitemOpen
  \bibfield  {author} {\bibinfo {author} {\bibfnamefont {O.}~\bibnamefont
  {{Kulakovich}}}, \bibinfo {author} {\bibfnamefont {N.}~\bibnamefont
  {{Strekal}}}, \bibinfo {author} {\bibfnamefont {A.}~\bibnamefont
  {{Yaroshevich}}}, \bibinfo {author} {\bibfnamefont {S.}~\bibnamefont
  {{Maskevich}}}, \bibinfo {author} {\bibfnamefont {S.}~\bibnamefont
  {{Gaponenko}}}, \bibinfo {author} {\bibfnamefont {I.}~\bibnamefont
  {{Nabiev}}}, \bibinfo {author} {\bibfnamefont {U.}~\bibnamefont {{Woggon}}},
  \ and\ \bibinfo {author} {\bibfnamefont {M.}~\bibnamefont {{Artemyev}}},\
  }\href {\doibase 10.1021/nl025819k} {\bibfield  {journal} {\bibinfo
  {journal} {Nano Lett.}\ }\textbf {\bibinfo {volume} {2}},\ \bibinfo {pages}
  {1449} (\bibinfo {year} {2002})}\BibitemShut {NoStop}%
\bibitem [{\citenamefont {Akselrod}\ \emph {et~al.}(2016)\citenamefont
  {Akselrod}, \citenamefont {Weidman}, \citenamefont {Li}, \citenamefont
  {Argyropoulos}, \citenamefont {Tisdale},\ and\ \citenamefont
  {Mikkelsen}}]{Akselrod2016}%
  \BibitemOpen
  \bibfield  {author} {\bibinfo {author} {\bibfnamefont {G.~M.}\ \bibnamefont
  {Akselrod}}, \bibinfo {author} {\bibfnamefont {M.~C.}\ \bibnamefont
  {Weidman}}, \bibinfo {author} {\bibfnamefont {Y.}~\bibnamefont {Li}},
  \bibinfo {author} {\bibfnamefont {C.}~\bibnamefont {Argyropoulos}}, \bibinfo
  {author} {\bibfnamefont {W.~A.}\ \bibnamefont {Tisdale}}, \ and\ \bibinfo
  {author} {\bibfnamefont {M.~H.}\ \bibnamefont {Mikkelsen}},\ }\href@noop {}
  {\bibfield  {journal} {\bibinfo  {journal} {ACS Photonics}\ }\textbf
  {\bibinfo {volume} {3}},\ \bibinfo {pages} {1741} (\bibinfo {year}
  {2016})}\BibitemShut {NoStop}%
\bibitem [{\citenamefont {David}\ \emph {et~al.}(2010)\citenamefont {David},
  \citenamefont {Richter}, \citenamefont {Knorr}, \citenamefont {Weidinger},\
  and\ \citenamefont {Hildebrandt}}]{david}%
  \BibitemOpen
  \bibfield  {author} {\bibinfo {author} {\bibfnamefont {C.}~\bibnamefont
  {David}}, \bibinfo {author} {\bibfnamefont {M.}~\bibnamefont {Richter}},
  \bibinfo {author} {\bibfnamefont {A.}~\bibnamefont {Knorr}}, \bibinfo
  {author} {\bibfnamefont {I.~M.}\ \bibnamefont {Weidinger}}, \ and\ \bibinfo
  {author} {\bibfnamefont {P.}~\bibnamefont {Hildebrandt}},\ }\href {\doibase
  10.1063/1.3291438} {\bibfield  {journal} {\bibinfo  {journal} {The Journal of
  Chemical Physics}\ }\textbf {\bibinfo {volume} {132}},\ \bibinfo {pages}
  {024712} (\bibinfo {year} {2010})}\BibitemShut {NoStop}%
\bibitem [{\citenamefont {Strelow}\ \emph {et~al.}(2016)\citenamefont
  {Strelow}, \citenamefont {Theuerholz}, \citenamefont {Schmidtke},
  \citenamefont {Richter}, \citenamefont {Merkl}, \citenamefont {Kloust},
  \citenamefont {Ye}, \citenamefont {Weller}, \citenamefont {Heinz},
  \citenamefont {Knorr},\ and\ \citenamefont {Lange}}]{NanoMarten}%
  \BibitemOpen
  \bibfield  {author} {\bibinfo {author} {\bibfnamefont {C.}~\bibnamefont
  {Strelow}}, \bibinfo {author} {\bibfnamefont {T.~S.}\ \bibnamefont
  {Theuerholz}}, \bibinfo {author} {\bibfnamefont {C.}~\bibnamefont
  {Schmidtke}}, \bibinfo {author} {\bibfnamefont {M.}~\bibnamefont {Richter}},
  \bibinfo {author} {\bibfnamefont {J.-P.}\ \bibnamefont {Merkl}}, \bibinfo
  {author} {\bibfnamefont {H.}~\bibnamefont {Kloust}}, \bibinfo {author}
  {\bibfnamefont {Z.}~\bibnamefont {Ye}}, \bibinfo {author} {\bibfnamefont
  {H.}~\bibnamefont {Weller}}, \bibinfo {author} {\bibfnamefont {T.~F.}\
  \bibnamefont {Heinz}}, \bibinfo {author} {\bibfnamefont {A.}~\bibnamefont
  {Knorr}}, \ and\ \bibinfo {author} {\bibfnamefont {H.}~\bibnamefont
  {Lange}},\ }\href@noop {} {\bibfield  {journal} {\bibinfo  {journal} {Nano
  Lett.}\ }\textbf {\bibinfo {volume} {16}},\ \bibinfo {pages} {4811} (\bibinfo
  {year} {2016})}\BibitemShut {NoStop}%
\bibitem [{\citenamefont {Barth}\ \emph {et~al.}(2010)\citenamefont {Barth},
  \citenamefont {Schietinger}, \citenamefont {Fischer}, \citenamefont {Becker},
  \citenamefont {N{\"{u}}sse}, \citenamefont {Aichele}, \citenamefont
  {L{\"{o}}chel}, \citenamefont {S{\"{o}}nnichsen},\ and\ \citenamefont
  {Benson}}]{Barth2010}%
  \BibitemOpen
  \bibfield  {author} {\bibinfo {author} {\bibfnamefont {M.}~\bibnamefont
  {Barth}}, \bibinfo {author} {\bibfnamefont {S.}~\bibnamefont {Schietinger}},
  \bibinfo {author} {\bibfnamefont {S.}~\bibnamefont {Fischer}}, \bibinfo
  {author} {\bibfnamefont {J.}~\bibnamefont {Becker}}, \bibinfo {author}
  {\bibfnamefont {N.}~\bibnamefont {N{\"{u}}sse}}, \bibinfo {author}
  {\bibfnamefont {T.}~\bibnamefont {Aichele}}, \bibinfo {author} {\bibfnamefont
  {B.}~\bibnamefont {L{\"{o}}chel}}, \bibinfo {author} {\bibfnamefont
  {C.}~\bibnamefont {S{\"{o}}nnichsen}}, \ and\ \bibinfo {author}
  {\bibfnamefont {O.}~\bibnamefont {Benson}},\ }\href {\doibase
  10.1021/nl903555u} {\bibfield  {journal} {\bibinfo  {journal} {Nano Letters}\
  }\textbf {\bibinfo {volume} {10}},\ \bibinfo {pages} {891} (\bibinfo {year}
  {2010})}\BibitemShut {NoStop}%
\bibitem [{\citenamefont {{Kamandar Dezfouli}}\ \emph
  {et~al.}(2017)\citenamefont {{Kamandar Dezfouli}}, \citenamefont {Gordon},\
  and\ \citenamefont {Hughes}}]{KamandarDezfouli2017}%
  \BibitemOpen
  \bibfield  {author} {\bibinfo {author} {\bibfnamefont {M.}~\bibnamefont
  {{Kamandar Dezfouli}}}, \bibinfo {author} {\bibfnamefont {R.}~\bibnamefont
  {Gordon}}, \ and\ \bibinfo {author} {\bibfnamefont {S.}~\bibnamefont
  {Hughes}},\ }\href {\doibase 10.1103/PhysRevA.95.013846} {\bibfield
  {journal} {\bibinfo  {journal} {Phys. Rev. A}\ }\textbf {\bibinfo {volume}
  {95}},\ \bibinfo {pages} {013846} (\bibinfo {year} {2017})}\BibitemShut
  {NoStop}%
\bibitem [{\citenamefont {Scully}\ \emph {et~al.}(1988)\citenamefont {Scully},
  \citenamefont {Wodkiewicz}, \citenamefont {Zubairy}, \citenamefont {Bergou},
  \citenamefont {Lu},\ and\ \citenamefont {ter Vehn}}]{lasing}%
  \BibitemOpen
  \bibfield  {author} {\bibinfo {author} {\bibfnamefont {M.~O.}\ \bibnamefont
  {Scully}}, \bibinfo {author} {\bibfnamefont {K.}~\bibnamefont {Wodkiewicz}},
  \bibinfo {author} {\bibfnamefont {M.}~\bibnamefont {Zubairy}}, \bibinfo
  {author} {\bibfnamefont {J.}~\bibnamefont {Bergou}}, \bibinfo {author}
  {\bibfnamefont {N.}~\bibnamefont {Lu}}, \ and\ \bibinfo {author}
  {\bibfnamefont {J.~M.}\ \bibnamefont {ter Vehn}},\ }\href@noop {} {\bibfield
  {journal} {\bibinfo  {journal} {Phys. Rev. Lett.}\ }\textbf {\bibinfo
  {volume} {60}},\ \bibinfo {pages} {1832} (\bibinfo {year}
  {1988})}\BibitemShut {NoStop}%
\bibitem [{\citenamefont {Bergman}\ and\ \citenamefont
  {Stockman}(2003)}]{spaser}%
  \BibitemOpen
  \bibfield  {author} {\bibinfo {author} {\bibfnamefont {D.~J.}\ \bibnamefont
  {Bergman}}\ and\ \bibinfo {author} {\bibfnamefont {M.~I.}\ \bibnamefont
  {Stockman}},\ }\href@noop {} {\bibfield  {journal} {\bibinfo  {journal}
  {Phys. Rev. Lett.}\ }\textbf {\bibinfo {volume} {90}},\ \bibinfo {pages}
  {027402} (\bibinfo {year} {2003})}\BibitemShut {NoStop}%
\bibitem [{\citenamefont {Kewes}\ \emph {et~al.}(2017)\citenamefont {Kewes},
  \citenamefont {Herrmann}, \citenamefont {Rodr\'{\i}guez-Oliveros},
  \citenamefont {Kuhlicke}, \citenamefont {Benson},\ and\ \citenamefont
  {Busch}}]{PhysRevLett.Spaser}%
  \BibitemOpen
  \bibfield  {author} {\bibinfo {author} {\bibfnamefont {G.}~\bibnamefont
  {Kewes}}, \bibinfo {author} {\bibfnamefont {K.}~\bibnamefont {Herrmann}},
  \bibinfo {author} {\bibfnamefont {R.}~\bibnamefont
  {Rodr\'{\i}guez-Oliveros}}, \bibinfo {author} {\bibfnamefont
  {A.}~\bibnamefont {Kuhlicke}}, \bibinfo {author} {\bibfnamefont
  {O.}~\bibnamefont {Benson}}, \ and\ \bibinfo {author} {\bibfnamefont
  {K.}~\bibnamefont {Busch}},\ }\href {\doibase 10.1103/PhysRevLett.118.237402}
  {\bibfield  {journal} {\bibinfo  {journal} {Phys. Rev. Lett.}\ }\textbf
  {\bibinfo {volume} {118}},\ \bibinfo {pages} {237402} (\bibinfo {year}
  {2017})}\BibitemShut {NoStop}%
\bibitem [{\citenamefont {Imamo\u{g}lu}\ \emph {et~al.}(1999)\citenamefont
  {Imamo\u{g}lu}, \citenamefont {Awschalom}, \citenamefont {Burkard},
  \citenamefont {DiVincenzo}, \citenamefont {Loss}, \citenamefont {Sherwin},\
  and\ \citenamefont {Small}}]{quantinfo}%
  \BibitemOpen
  \bibfield  {author} {\bibinfo {author} {\bibfnamefont {A.}~\bibnamefont
  {Imamo\u{g}lu}}, \bibinfo {author} {\bibfnamefont {D.~D.}\ \bibnamefont
  {Awschalom}}, \bibinfo {author} {\bibfnamefont {G.}~\bibnamefont {Burkard}},
  \bibinfo {author} {\bibfnamefont {D.~P.}\ \bibnamefont {DiVincenzo}},
  \bibinfo {author} {\bibfnamefont {D.}~\bibnamefont {Loss}}, \bibinfo {author}
  {\bibfnamefont {M.}~\bibnamefont {Sherwin}}, \ and\ \bibinfo {author}
  {\bibfnamefont {A.}~\bibnamefont {Small}},\ }\href {\doibase
  10.1103/PhysRevLett.83.4204} {\bibfield  {journal} {\bibinfo  {journal}
  {Phys. Rev. Lett.}\ }\textbf {\bibinfo {volume} {83}},\ \bibinfo {pages}
  {4204} (\bibinfo {year} {1999})}\BibitemShut {NoStop}%
\bibitem [{\citenamefont {Gea-Banacloche}\ \emph {et~al.}(1990)\citenamefont
  {Gea-Banacloche}, \citenamefont {Lu}, \citenamefont {Pedrotti}, \citenamefont
  {Prasad}, \citenamefont {Scully},\ and\ \citenamefont
  {W{\'o}dkiewicz}}]{motu1}%
  \BibitemOpen
  \bibfield  {author} {\bibinfo {author} {\bibfnamefont {J.}~\bibnamefont
  {Gea-Banacloche}}, \bibinfo {author} {\bibfnamefont {N.}~\bibnamefont {Lu}},
  \bibinfo {author} {\bibfnamefont {L.~M.}\ \bibnamefont {Pedrotti}}, \bibinfo
  {author} {\bibfnamefont {S.}~\bibnamefont {Prasad}}, \bibinfo {author}
  {\bibfnamefont {M.~O.}\ \bibnamefont {Scully}}, \ and\ \bibinfo {author}
  {\bibfnamefont {K.}~\bibnamefont {W{\'o}dkiewicz}},\ }\href@noop {}
  {\bibfield  {journal} {\bibinfo  {journal} {Phys. Rev. A}\ }\textbf {\bibinfo
  {volume} {41}},\ \bibinfo {pages} {369} (\bibinfo {year} {1990})}\BibitemShut
  {NoStop}%
\bibitem [{\citenamefont {Glauber}\ and\ \citenamefont
  {Lewenstein}(1991)}]{motu2}%
  \BibitemOpen
  \bibfield  {author} {\bibinfo {author} {\bibfnamefont {R.~J.}\ \bibnamefont
  {Glauber}}\ and\ \bibinfo {author} {\bibfnamefont {M.}~\bibnamefont
  {Lewenstein}},\ }\href {\doibase 10.1103/PhysRevA.43.467} {\bibfield
  {journal} {\bibinfo  {journal} {Phys. Rev. A}\ }\textbf {\bibinfo {volume}
  {43}},\ \bibinfo {pages} {467} (\bibinfo {year} {1991})}\BibitemShut
  {NoStop}%
\bibitem [{\citenamefont {Dalton}\ \emph {et~al.}(1996)\citenamefont {Dalton},
  \citenamefont {Guerra},\ and\ \citenamefont {Knight}}]{motu3}%
  \BibitemOpen
  \bibfield  {author} {\bibinfo {author} {\bibfnamefont {B.}~\bibnamefont
  {Dalton}}, \bibinfo {author} {\bibfnamefont {E.}~\bibnamefont {Guerra}}, \
  and\ \bibinfo {author} {\bibfnamefont {P.}~\bibnamefont {Knight}},\
  }\href@noop {} {\bibfield  {journal} {\bibinfo  {journal} {Phys. Rev. A}\
  }\textbf {\bibinfo {volume} {54}},\ \bibinfo {pages} {2292} (\bibinfo {year}
  {1996})}\BibitemShut {NoStop}%
\bibitem [{\citenamefont {Kweon}\ and\ \citenamefont {Lawandy}(1995)}]{motu4}%
  \BibitemOpen
  \bibfield  {author} {\bibinfo {author} {\bibfnamefont {G.-I.}\ \bibnamefont
  {Kweon}}\ and\ \bibinfo {author} {\bibfnamefont {N.}~\bibnamefont
  {Lawandy}},\ }\href@noop {} {\bibfield  {journal} {\bibinfo  {journal} {Opt.
  Commun.}\ }\textbf {\bibinfo {volume} {118}},\ \bibinfo {pages} {388}
  (\bibinfo {year} {1995})}\BibitemShut {NoStop}%
\bibitem [{\citenamefont {Jaynes}\ and\ \citenamefont
  {Cummings}(1963)}]{Jaynes_ProcIEEE_51_89_1963}%
  \BibitemOpen
  \bibfield  {author} {\bibinfo {author} {\bibfnamefont {E.~T.}\ \bibnamefont
  {Jaynes}}\ and\ \bibinfo {author} {\bibfnamefont {F.~W.}\ \bibnamefont
  {Cummings}},\ }\href@noop {} {\bibfield  {journal} {\bibinfo  {journal}
  {Proceedings of the IEEE}\ }\textbf {\bibinfo {volume} {51}},\ \bibinfo
  {pages} {89} (\bibinfo {year} {1963})}\BibitemShut {NoStop}%
\bibitem [{\citenamefont {Agarwal}(2013)}]{GirishBook1}%
  \BibitemOpen
  \bibfield  {author} {\bibinfo {author} {\bibfnamefont {G.~S.}\ \bibnamefont
  {Agarwal}},\ }\href@noop {} {\emph {\bibinfo {title} {Quantum Optics}}}\
  (\bibinfo  {publisher} {Cambridge University Press},\ \bibinfo {year}
  {2013})\BibitemShut {NoStop}%
\bibitem [{\citenamefont {Zrenner}\ \emph {et~al.}(2002)\citenamefont
  {Zrenner}, \citenamefont {Beham}, \citenamefont {Stufler}, \citenamefont
  {Findeis}, \citenamefont {Bichler},\ and\ \citenamefont
  {Abstreiter}}]{2levelsys1}%
  \BibitemOpen
  \bibfield  {author} {\bibinfo {author} {\bibfnamefont {A.}~\bibnamefont
  {Zrenner}}, \bibinfo {author} {\bibfnamefont {E.}~\bibnamefont {Beham}},
  \bibinfo {author} {\bibfnamefont {S.}~\bibnamefont {Stufler}}, \bibinfo
  {author} {\bibfnamefont {F.}~\bibnamefont {Findeis}}, \bibinfo {author}
  {\bibfnamefont {M.}~\bibnamefont {Bichler}}, \ and\ \bibinfo {author}
  {\bibfnamefont {G.}~\bibnamefont {Abstreiter}},\ }\href@noop {} {\bibfield
  {journal} {\bibinfo  {journal} {Nature}\ }\textbf {\bibinfo {volume} {418}},\
  \bibinfo {pages} {612} (\bibinfo {year} {2002})}\BibitemShut {NoStop}%
\bibitem [{\citenamefont {Gegg}\ and\ \citenamefont {Richter}(2016)}]{gegg}%
  \BibitemOpen
  \bibfield  {author} {\bibinfo {author} {\bibfnamefont {M.}~\bibnamefont
  {Gegg}}\ and\ \bibinfo {author} {\bibfnamefont {M.}~\bibnamefont {Richter}},\
  }\href@noop {} {\bibfield  {journal} {\bibinfo  {journal} {New J. Phys.}\
  }\textbf {\bibinfo {volume} {18}},\ \bibinfo {pages} {043037} (\bibinfo
  {year} {2016})}\BibitemShut {NoStop}%
\bibitem [{\citenamefont {Lodahl}\ \emph {et~al.}(2015)\citenamefont {Lodahl},
  \citenamefont {Mahmoodian},\ and\ \citenamefont
  {Stobbe}}]{RevModPhys.87.347}%
  \BibitemOpen
  \bibfield  {author} {\bibinfo {author} {\bibfnamefont {P.}~\bibnamefont
  {Lodahl}}, \bibinfo {author} {\bibfnamefont {S.}~\bibnamefont {Mahmoodian}},
  \ and\ \bibinfo {author} {\bibfnamefont {S.}~\bibnamefont {Stobbe}},\ }\href
  {\doibase 10.1103/RevModPhys.87.347} {\bibfield  {journal} {\bibinfo
  {journal} {Rev. Mod. Phys.}\ }\textbf {\bibinfo {volume} {87}},\ \bibinfo
  {pages} {347} (\bibinfo {year} {2015})}\BibitemShut {NoStop}%
\bibitem [{\citenamefont {Gardiner}\ and\ \citenamefont
  {Zoller}(2004)}]{QNoise}%
  \BibitemOpen
  \bibfield  {author} {\bibinfo {author} {\bibfnamefont {C.~W.}\ \bibnamefont
  {Gardiner}}\ and\ \bibinfo {author} {\bibfnamefont {P.}~\bibnamefont
  {Zoller}},\ }\href@noop {} {\emph {\bibinfo {title} {Quantum Noise}}}\
  (\bibinfo  {publisher} {Springer},\ \bibinfo {year} {2004})\BibitemShut
  {NoStop}%
\bibitem [{\citenamefont {Gardiner}\ and\ \citenamefont
  {Collett}(1985)}]{Gardiner1}%
  \BibitemOpen
  \bibfield  {author} {\bibinfo {author} {\bibfnamefont {C.~W.}\ \bibnamefont
  {Gardiner}}\ and\ \bibinfo {author} {\bibfnamefont {M.~J.}\ \bibnamefont
  {Collett}},\ }\href {\doibase 10.1103/PhysRevA.31.3761} {\bibfield  {journal}
  {\bibinfo  {journal} {Phys. Rev. A}\ }\textbf {\bibinfo {volume} {31}},\
  \bibinfo {pages} {3761} (\bibinfo {year} {1985})}\BibitemShut {NoStop}%
\bibitem [{\citenamefont {Waks}\ and\ \citenamefont
  {Sridharan}(2010)}]{PhysRevA.82.043845}%
  \BibitemOpen
  \bibfield  {author} {\bibinfo {author} {\bibfnamefont {E.}~\bibnamefont
  {Waks}}\ and\ \bibinfo {author} {\bibfnamefont {D.}~\bibnamefont
  {Sridharan}},\ }\href {\doibase 10.1103/PhysRevA.82.043845} {\bibfield
  {journal} {\bibinfo  {journal} {Phys. Rev. A}\ }\textbf {\bibinfo {volume}
  {82}},\ \bibinfo {pages} {043845} (\bibinfo {year} {2010})}\BibitemShut
  {NoStop}%
\bibitem [{\citenamefont {Dignam}\ and\ \citenamefont
  {Dezfouli}(2012)}]{Dignam}%
  \BibitemOpen
  \bibfield  {author} {\bibinfo {author} {\bibfnamefont {M.~M.}\ \bibnamefont
  {Dignam}}\ and\ \bibinfo {author} {\bibfnamefont {M.~K.}\ \bibnamefont
  {Dezfouli}},\ }\href {\doibase 10.1103/PhysRevA.85.013809} {\bibfield
  {journal} {\bibinfo  {journal} {Phys. Rev. A}\ }\textbf {\bibinfo {volume}
  {85}},\ \bibinfo {pages} {013809} (\bibinfo {year} {2012})}\BibitemShut
  {NoStop}%
\bibitem [{\citenamefont {Hughes}\ \emph {et~al.}(2018)\citenamefont {Hughes},
  \citenamefont {Richter},\ and\ \citenamefont {Knorr}}]{Hughes:18}%
  \BibitemOpen
  \bibfield  {author} {\bibinfo {author} {\bibfnamefont {S.}~\bibnamefont
  {Hughes}}, \bibinfo {author} {\bibfnamefont {M.}~\bibnamefont {Richter}}, \
  and\ \bibinfo {author} {\bibfnamefont {A.}~\bibnamefont {Knorr}},\ }\href
  {\doibase 10.1364/OL.43.001834} {\bibfield  {journal} {\bibinfo  {journal}
  {Opt. Lett.}\ }\textbf {\bibinfo {volume} {43}},\ \bibinfo {pages} {1834}
  (\bibinfo {year} {2018})}\BibitemShut {NoStop}%
\bibitem [{\citenamefont {Hensen}\ \emph {et~al.}(2017)\citenamefont {Hensen},
  \citenamefont {Heilpern}, \citenamefont {Gray},\ and\ \citenamefont
  {Pfeiffer}}]{hensenstrong}%
  \BibitemOpen
  \bibfield  {author} {\bibinfo {author} {\bibfnamefont {M.}~\bibnamefont
  {Hensen}}, \bibinfo {author} {\bibfnamefont {T.}~\bibnamefont {Heilpern}},
  \bibinfo {author} {\bibfnamefont {S.~K.}\ \bibnamefont {Gray}}, \ and\
  \bibinfo {author} {\bibfnamefont {W.}~\bibnamefont {Pfeiffer}},\ }\href@noop
  {} {\bibfield  {journal} {\bibinfo  {journal} {ACS photonics}\ }\textbf
  {\bibinfo {volume} {5}},\ \bibinfo {pages} {240} (\bibinfo {year}
  {2017})}\BibitemShut {NoStop}%
\bibitem [{\citenamefont {Ge}\ and\ \citenamefont
  {Hughes}(2015)}]{PhysRevB.92.205420}%
  \BibitemOpen
  \bibfield  {author} {\bibinfo {author} {\bibfnamefont {R.-C.}\ \bibnamefont
  {Ge}}\ and\ \bibinfo {author} {\bibfnamefont {S.}~\bibnamefont {Hughes}},\
  }\href {\doibase 10.1103/PhysRevB.92.205420} {\bibfield  {journal} {\bibinfo
  {journal} {Phys. Rev. B}\ }\textbf {\bibinfo {volume} {92}},\ \bibinfo
  {pages} {205420} (\bibinfo {year} {2015})}\BibitemShut {NoStop}%
\bibitem [{\citenamefont {Lee}\ \emph {et~al.}(1999)\citenamefont {Lee},
  \citenamefont {Leung},\ and\ \citenamefont {Pang}}]{MDR1}%
  \BibitemOpen
  \bibfield  {author} {\bibinfo {author} {\bibfnamefont {K.~M.}\ \bibnamefont
  {Lee}}, \bibinfo {author} {\bibfnamefont {P.~T.}\ \bibnamefont {Leung}}, \
  and\ \bibinfo {author} {\bibfnamefont {K.~M.}\ \bibnamefont {Pang}},\ }\href
  {\doibase 10.1364/JOSAB.16.001409} {\bibfield  {journal} {\bibinfo  {journal}
  {J. Opt. Soc. Am. B}\ }\textbf {\bibinfo {volume} {16}},\ \bibinfo {pages}
  {1409} (\bibinfo {year} {1999})}\BibitemShut {NoStop}%
\bibitem [{\citenamefont {Muljarov}\ \emph {et~al.}(2010)\citenamefont
  {Muljarov}, \citenamefont {Langbein},\ and\ \citenamefont
  {Zimmermann}}]{muljarovPert}%
  \BibitemOpen
  \bibfield  {author} {\bibinfo {author} {\bibfnamefont {E.~A.}\ \bibnamefont
  {Muljarov}}, \bibinfo {author} {\bibfnamefont {W.}~\bibnamefont {Langbein}},
  \ and\ \bibinfo {author} {\bibfnamefont {R.}~\bibnamefont {Zimmermann}},\
  }\href@noop {} {\bibfield  {journal} {\bibinfo  {journal} {EPL (Europhysics
  Letters)}\ }\textbf {\bibinfo {volume} {92}},\ \bibinfo {pages} {50010}
  (\bibinfo {year} {2010})}\BibitemShut {NoStop}%
\bibitem [{\citenamefont {Kristensen}\ \emph {et~al.}(2012)\citenamefont
  {Kristensen}, \citenamefont {Van~Vlack},\ and\ \citenamefont
  {Hughes}}]{KristensenHughes}%
  \BibitemOpen
  \bibfield  {author} {\bibinfo {author} {\bibfnamefont {P.~T.}\ \bibnamefont
  {Kristensen}}, \bibinfo {author} {\bibfnamefont {C.}~\bibnamefont
  {Van~Vlack}}, \ and\ \bibinfo {author} {\bibfnamefont {S.}~\bibnamefont
  {Hughes}},\ }\href {\doibase 10.1364/OL.37.001649} {\bibfield  {journal}
  {\bibinfo  {journal} {Opt. Lett.}\ }\textbf {\bibinfo {volume} {37}},\
  \bibinfo {pages} {1649} (\bibinfo {year} {2012})}\BibitemShut {NoStop}%
\bibitem [{\citenamefont {Sauvan}\ \emph {et~al.}(2013)\citenamefont {Sauvan},
  \citenamefont {Hugonin}, \citenamefont {Maksymov},\ and\ \citenamefont
  {Lalanne}}]{SauvanNorm}%
  \BibitemOpen
  \bibfield  {author} {\bibinfo {author} {\bibfnamefont {C.}~\bibnamefont
  {Sauvan}}, \bibinfo {author} {\bibfnamefont {J.~P.}\ \bibnamefont {Hugonin}},
  \bibinfo {author} {\bibfnamefont {I.~S.}\ \bibnamefont {Maksymov}}, \ and\
  \bibinfo {author} {\bibfnamefont {P.}~\bibnamefont {Lalanne}},\ }\href
  {\doibase 10.1103/PhysRevLett.110.237401} {\bibfield  {journal} {\bibinfo
  {journal} {Phys. Rev. Lett.}\ }\textbf {\bibinfo {volume} {110}},\ \bibinfo
  {pages} {237401} (\bibinfo {year} {2013})}\BibitemShut {NoStop}%
\bibitem [{\citenamefont {Kristensen}\ and\ \citenamefont
  {Hughes}(2014)}]{NormKristHughes}%
  \BibitemOpen
  \bibfield  {author} {\bibinfo {author} {\bibfnamefont {P.~T.}\ \bibnamefont
  {Kristensen}}\ and\ \bibinfo {author} {\bibfnamefont {S.}~\bibnamefont
  {Hughes}},\ }\href@noop {} {\bibfield  {journal} {\bibinfo  {journal} {ACS
  Photonics}\ }\textbf {\bibinfo {volume} {1}},\ \bibinfo {pages} {2} (\bibinfo
  {year} {2014})}\BibitemShut {NoStop}%
\bibitem [{\citenamefont {Lalanne}\ \emph {et~al.}(2018)\citenamefont
  {Lalanne}, \citenamefont {Yan}, \citenamefont {Vynck}, \citenamefont
  {Sauvan},\ and\ \citenamefont {Hugonin}}]{Lalanne_review}%
  \BibitemOpen
  \bibfield  {author} {\bibinfo {author} {\bibfnamefont {P.}~\bibnamefont
  {Lalanne}}, \bibinfo {author} {\bibfnamefont {W.}~\bibnamefont {Yan}},
  \bibinfo {author} {\bibfnamefont {K.}~\bibnamefont {Vynck}}, \bibinfo
  {author} {\bibfnamefont {C.}~\bibnamefont {Sauvan}}, \ and\ \bibinfo {author}
  {\bibfnamefont {J.-P.}\ \bibnamefont {Hugonin}},\ }\href@noop {} {\bibfield
  {journal} {\bibinfo  {journal} {Laser \& Photonics Reviews}\ }\textbf
  {\bibinfo {volume} {12}},\ \bibinfo {pages} {1700113} (\bibinfo {year}
  {2018})}\BibitemShut {NoStop}%
\bibitem [{\citenamefont {{Fern\'{a}ndez-Dom\'{i}nguez, Antonio I. and
  Bozhevolnyi, Sergey I. and Mortensen, N. Asger}}(2018)}]{ACSAsger}%
  \BibitemOpen
  \bibfield  {author} {\bibinfo {author} {\bibnamefont
  {{Fern\'{a}ndez-Dom\'{i}nguez, Antonio I. and Bozhevolnyi, Sergey I. and
  Mortensen, N. Asger}}},\ }\href {\doibase 10.1021/acsphotonics.8b00852}
  {\bibfield  {journal} {\bibinfo  {journal} {ACS Photonics}\ }\textbf
  {\bibinfo {volume} {5}},\ \bibinfo {pages} {3447} (\bibinfo {year}
  {2018})}\BibitemShut {NoStop}%
\bibitem [{\citenamefont {Ho}\ \emph {et~al.}(1998)\citenamefont {Ho},
  \citenamefont {Leung}, \citenamefont {Maassen van~den Brink},\ and\
  \citenamefont {Young}}]{2ndquanho}%
  \BibitemOpen
  \bibfield  {author} {\bibinfo {author} {\bibfnamefont {K.~C.}\ \bibnamefont
  {Ho}}, \bibinfo {author} {\bibfnamefont {P.~T.}\ \bibnamefont {Leung}},
  \bibinfo {author} {\bibfnamefont {A.}~\bibnamefont {Maassen van~den Brink}},
  \ and\ \bibinfo {author} {\bibfnamefont {K.}~\bibnamefont {Young}},\ }\href
  {\doibase 10.1103/PhysRevE.58.2965} {\bibfield  {journal} {\bibinfo
  {journal} {Phys. Rev. E}\ }\textbf {\bibinfo {volume} {58}},\ \bibinfo
  {pages} {2965} (\bibinfo {year} {1998})}\BibitemShut {NoStop}%
\bibitem [{\citenamefont {Severini}\ \emph {et~al.}(2004)\citenamefont
  {Severini}, \citenamefont {Settimi}, \citenamefont {Sibilia}, \citenamefont
  {Bertolotti}, \citenamefont {Napoli},\ and\ \citenamefont
  {Messina}}]{Severini}%
  \BibitemOpen
  \bibfield  {author} {\bibinfo {author} {\bibfnamefont {S.}~\bibnamefont
  {Severini}}, \bibinfo {author} {\bibfnamefont {A.}~\bibnamefont {Settimi}},
  \bibinfo {author} {\bibfnamefont {C.}~\bibnamefont {Sibilia}}, \bibinfo
  {author} {\bibfnamefont {M.}~\bibnamefont {Bertolotti}}, \bibinfo {author}
  {\bibfnamefont {A.}~\bibnamefont {Napoli}}, \ and\ \bibinfo {author}
  {\bibfnamefont {A.}~\bibnamefont {Messina}},\ }\href {\doibase
  10.1103/PhysRevE.70.056614} {\bibfield  {journal} {\bibinfo  {journal} {Phys.
  Rev. E}\ }\textbf {\bibinfo {volume} {70}},\ \bibinfo {pages} {056614}
  (\bibinfo {year} {2004})}\BibitemShut {NoStop}%
\bibitem [{\citenamefont {Huttner}\ and\ \citenamefont
  {Barnett}(1992)}]{HuttnerBarnett}%
  \BibitemOpen
  \bibfield  {author} {\bibinfo {author} {\bibfnamefont {B.}~\bibnamefont
  {Huttner}}\ and\ \bibinfo {author} {\bibfnamefont {S.~M.}\ \bibnamefont
  {Barnett}},\ }\href {\doibase 10.1103/PhysRevA.46.4306} {\bibfield  {journal}
  {\bibinfo  {journal} {Phys. Rev. A}\ }\textbf {\bibinfo {volume} {46}},\
  \bibinfo {pages} {4306} (\bibinfo {year} {1992})}\BibitemShut {NoStop}%
\bibitem [{\citenamefont {Tip}(1997)}]{A.Tip}%
  \BibitemOpen
  \bibfield  {author} {\bibinfo {author} {\bibfnamefont {A.}~\bibnamefont
  {Tip}},\ }\href {\doibase 10.1103/PhysRevA.56.5022} {\bibfield  {journal}
  {\bibinfo  {journal} {Phys. Rev. A}\ }\textbf {\bibinfo {volume} {56}},\
  \bibinfo {pages} {5022} (\bibinfo {year} {1997})}\BibitemShut {NoStop}%
\bibitem [{\citenamefont {Dung}\ \emph {et~al.}(1998)\citenamefont {Dung},
  \citenamefont {Kn\"oll},\ and\ \citenamefont {Welsch}}]{Dung}%
  \BibitemOpen
  \bibfield  {author} {\bibinfo {author} {\bibfnamefont {H.~T.}\ \bibnamefont
  {Dung}}, \bibinfo {author} {\bibfnamefont {L.}~\bibnamefont {Kn\"oll}}, \
  and\ \bibinfo {author} {\bibfnamefont {D.-G.}\ \bibnamefont {Welsch}},\
  }\href {\doibase 10.1103/PhysRevA.57.3931} {\bibfield  {journal} {\bibinfo
  {journal} {Phys. Rev. A}\ }\textbf {\bibinfo {volume} {57}},\ \bibinfo
  {pages} {3931} (\bibinfo {year} {1998})}\BibitemShut {NoStop}%
\bibitem [{\citenamefont {Suttorp}\ and\ \citenamefont {van
  Wonderen}(2004)}]{Suttorp}%
  \BibitemOpen
  \bibfield  {author} {\bibinfo {author} {\bibfnamefont {L.~G.}\ \bibnamefont
  {Suttorp}}\ and\ \bibinfo {author} {\bibfnamefont {A.~J.}\ \bibnamefont {van
  Wonderen}},\ }\href {http://stacks.iop.org/0295-5075/67/i=5/a=766} {\bibfield
   {journal} {\bibinfo  {journal} {EPL (Europhysics Letters)}\ }\textbf
  {\bibinfo {volume} {67}},\ \bibinfo {pages} {766} (\bibinfo {year}
  {2004})}\BibitemShut {NoStop}%
\bibitem [{\citenamefont {Gruner}\ and\ \citenamefont
  {Welsch}(1996)}]{grunwel}%
  \BibitemOpen
  \bibfield  {author} {\bibinfo {author} {\bibfnamefont {T.}~\bibnamefont
  {Gruner}}\ and\ \bibinfo {author} {\bibfnamefont {D.-G.}\ \bibnamefont
  {Welsch}},\ }\href {\doibase 10.1103/PhysRevA.53.1818} {\bibfield  {journal}
  {\bibinfo  {journal} {Phys. Rev. A}\ }\textbf {\bibinfo {volume} {53}},\
  \bibinfo {pages} {1818} (\bibinfo {year} {1996})}\BibitemShut {NoStop}%
\bibitem [{\citenamefont {Kristensen}\ \emph {et~al.}(2015)\citenamefont
  {Kristensen}, \citenamefont {Ge},\ and\ \citenamefont {Hughes}}]{normaliz}%
  \BibitemOpen
  \bibfield  {author} {\bibinfo {author} {\bibfnamefont {P.~T.}\ \bibnamefont
  {Kristensen}}, \bibinfo {author} {\bibfnamefont {R.-C.}\ \bibnamefont {Ge}},
  \ and\ \bibinfo {author} {\bibfnamefont {S.}~\bibnamefont {Hughes}},\ }\href
  {\doibase 10.1103/PhysRevA.92.053810} {\bibfield  {journal} {\bibinfo
  {journal} {Phys. Rev. A}\ }\textbf {\bibinfo {volume} {92}},\ \bibinfo
  {pages} {053810} (\bibinfo {year} {2015})}\BibitemShut {NoStop}%
\bibitem [{\citenamefont {Doost}\ \emph {et~al.}(2013)\citenamefont {Doost},
  \citenamefont {Langbein},\ and\ \citenamefont
  {Muljarov}}]{Doost_PRA_87_043827_2013}%
  \BibitemOpen
  \bibfield  {author} {\bibinfo {author} {\bibfnamefont {M.~B.}\ \bibnamefont
  {Doost}}, \bibinfo {author} {\bibfnamefont {W.}~\bibnamefont {Langbein}}, \
  and\ \bibinfo {author} {\bibfnamefont {E.~A.}\ \bibnamefont {Muljarov}},\
  }\href@noop {} {\bibfield  {journal} {\bibinfo  {journal} {Phys. Rev. A}\
  }\textbf {\bibinfo {volume} {87}},\ \bibinfo {pages} {043827} (\bibinfo
  {year} {2013})}\BibitemShut {NoStop}%
\bibitem [{\citenamefont {Ge}\ \emph {et~al.}(2014)\citenamefont {Ge},
  \citenamefont {Kristensen}, \citenamefont {Young},\ and\ \citenamefont
  {Hughes}}]{Ge1}%
  \BibitemOpen
  \bibfield  {author} {\bibinfo {author} {\bibfnamefont {R.-C.}\ \bibnamefont
  {Ge}}, \bibinfo {author} {\bibfnamefont {P.~T.}\ \bibnamefont {Kristensen}},
  \bibinfo {author} {\bibfnamefont {J.~F.}\ \bibnamefont {Young}}, \ and\
  \bibinfo {author} {\bibfnamefont {S.}~\bibnamefont {Hughes}},\ }\href
  {http://stacks.iop.org/1367-2630/16/i=11/a=113048} {\bibfield  {journal}
  {\bibinfo  {journal} {New J. Phys.}\ }\textbf {\bibinfo {volume} {16}},\
  \bibinfo {pages} {113048} (\bibinfo {year} {2014})}\BibitemShut {NoStop}%
\bibitem [{\citenamefont {H\"ummer}\ \emph {et~al.}(2013)\citenamefont
  {H\"ummer}, \citenamefont {Garc\'{\i}a-Vidal}, \citenamefont
  {Mart\'{\i}n-Moreno},\ and\ \citenamefont {Zueco}}]{PlasQED}%
  \BibitemOpen
  \bibfield  {author} {\bibinfo {author} {\bibfnamefont {T.}~\bibnamefont
  {H\"ummer}}, \bibinfo {author} {\bibfnamefont {F.~J.}\ \bibnamefont
  {Garc\'{\i}a-Vidal}}, \bibinfo {author} {\bibfnamefont {L.}~\bibnamefont
  {Mart\'{\i}n-Moreno}}, \ and\ \bibinfo {author} {\bibfnamefont
  {D.}~\bibnamefont {Zueco}},\ }\href {\doibase 10.1103/PhysRevB.87.115419}
  {\bibfield  {journal} {\bibinfo  {journal} {Phys. Rev. B}\ }\textbf {\bibinfo
  {volume} {87}},\ \bibinfo {pages} {115419} (\bibinfo {year}
  {2013})}\BibitemShut {NoStop}%
\bibitem [{\citenamefont {Varguet}\ \emph {et~al.}(2016)\citenamefont
  {Varguet}, \citenamefont {Rousseaux}, \citenamefont {Dzsotjan}, \citenamefont
  {Jauslin}, \citenamefont {Gu\'{e}rin},\ and\ \citenamefont {des
  Francs}}]{Rouss1}%
  \BibitemOpen
  \bibfield  {author} {\bibinfo {author} {\bibfnamefont {H.}~\bibnamefont
  {Varguet}}, \bibinfo {author} {\bibfnamefont {B.}~\bibnamefont {Rousseaux}},
  \bibinfo {author} {\bibfnamefont {D.}~\bibnamefont {Dzsotjan}}, \bibinfo
  {author} {\bibfnamefont {H.~R.}\ \bibnamefont {Jauslin}}, \bibinfo {author}
  {\bibfnamefont {S.}~\bibnamefont {Gu\'{e}rin}}, \ and\ \bibinfo {author}
  {\bibfnamefont {G.~C.}\ \bibnamefont {des Francs}},\ }\href {\doibase
  10.1364/OL.41.004480} {\bibfield  {journal} {\bibinfo  {journal} {Opt.
  Lett.}\ }\textbf {\bibinfo {volume} {41}},\ \bibinfo {pages} {4480} (\bibinfo
  {year} {2016})}\BibitemShut {NoStop}%
\bibitem [{\citenamefont {Dzsotjan}\ \emph {et~al.}(2016)\citenamefont
  {Dzsotjan}, \citenamefont {Rousseaux}, \citenamefont {Jauslin}, \citenamefont
  {des Francs}, \citenamefont {Couteau},\ and\ \citenamefont
  {Gu\'erin}}]{Rouss2}%
  \BibitemOpen
  \bibfield  {author} {\bibinfo {author} {\bibfnamefont {D.}~\bibnamefont
  {Dzsotjan}}, \bibinfo {author} {\bibfnamefont {B.}~\bibnamefont {Rousseaux}},
  \bibinfo {author} {\bibfnamefont {H.~R.}\ \bibnamefont {Jauslin}}, \bibinfo
  {author} {\bibfnamefont {G.~C.}\ \bibnamefont {des Francs}}, \bibinfo
  {author} {\bibfnamefont {C.}~\bibnamefont {Couteau}}, \ and\ \bibinfo
  {author} {\bibfnamefont {S.}~\bibnamefont {Gu\'erin}},\ }\href {\doibase
  10.1103/PhysRevA.94.023818} {\bibfield  {journal} {\bibinfo  {journal} {Phys.
  Rev. A}\ }\textbf {\bibinfo {volume} {94}},\ \bibinfo {pages} {023818}
  (\bibinfo {year} {2016})}\BibitemShut {NoStop}%
\bibitem [{\citenamefont {Rousseaux}\ \emph {et~al.}(2016)\citenamefont
  {Rousseaux}, \citenamefont {Dzsotjan}, \citenamefont {Colas~des Francs},
  \citenamefont {Jauslin}, \citenamefont {Couteau},\ and\ \citenamefont
  {Gu\'erin}}]{Rouss3}%
  \BibitemOpen
  \bibfield  {author} {\bibinfo {author} {\bibfnamefont {B.}~\bibnamefont
  {Rousseaux}}, \bibinfo {author} {\bibfnamefont {D.}~\bibnamefont {Dzsotjan}},
  \bibinfo {author} {\bibfnamefont {G.}~\bibnamefont {Colas~des Francs}},
  \bibinfo {author} {\bibfnamefont {H.~R.}\ \bibnamefont {Jauslin}}, \bibinfo
  {author} {\bibfnamefont {C.}~\bibnamefont {Couteau}}, \ and\ \bibinfo
  {author} {\bibfnamefont {S.}~\bibnamefont {Gu\'erin}},\ }\href {\doibase
  10.1103/PhysRevB.93.045422} {\bibfield  {journal} {\bibinfo  {journal} {Phys.
  Rev. B}\ }\textbf {\bibinfo {volume} {93}},\ \bibinfo {pages} {045422}
  (\bibinfo {year} {2016})}\BibitemShut {NoStop}%
\bibitem [{\citenamefont {Ge}\ and\ \citenamefont {Hughes}(2014)}]{Ge:14Dimer}%
  \BibitemOpen
  \bibfield  {author} {\bibinfo {author} {\bibfnamefont {R.-C.}\ \bibnamefont
  {Ge}}\ and\ \bibinfo {author} {\bibfnamefont {S.}~\bibnamefont {Hughes}},\
  }\href {\doibase 10.1364/OL.39.004235} {\bibfield  {journal} {\bibinfo
  {journal} {Opt. Lett.}\ }\textbf {\bibinfo {volume} {39}},\ \bibinfo {pages}
  {4235} (\bibinfo {year} {2014})}\BibitemShut {NoStop}%
\bibitem [{SI()}]{SI}%
  \BibitemOpen
  \href@noop {} {}\bibinfo {note} {See Supplemental Material at [URL will be
  inserted by publisher] for detailed calculations, including the derivation of
  the commutation relations, the Heisenberg equation of motion of the
  quasinormal mode operators, the derivation of the Lindblad master equation,
  the bad cavity limit for the multi-QNM case as well as a discussion about
  details of the numerical calculations}\BibitemShut {NoStop}%
\bibitem [{\citenamefont {L\"owdin}(1950)}]{Chemie2}%
  \BibitemOpen
  \bibfield  {author} {\bibinfo {author} {\bibfnamefont {P.}~\bibnamefont
  {L\"owdin}},\ }\href@noop {} {\bibfield  {journal} {\bibinfo  {journal} {J.
  Chem. Phys}\ }\textbf {\bibinfo {volume} {18}},\ \bibinfo {pages} {365}
  (\bibinfo {year} {1950})}\BibitemShut {NoStop}%
\bibitem [{\citenamefont {Lax}(1966)}]{Lax}%
  \BibitemOpen
  \bibfield  {author} {\bibinfo {author} {\bibfnamefont {M.}~\bibnamefont
  {Lax}},\ }\href {\doibase 10.1103/PhysRev.145.110} {\bibfield  {journal}
  {\bibinfo  {journal} {Phys. Rev.}\ }\textbf {\bibinfo {volume} {145}},\
  \bibinfo {pages} {110} (\bibinfo {year} {1966})}\BibitemShut {NoStop}%
\bibitem [{\citenamefont {de~Lasson}\ \emph {et~al.}(2015)\citenamefont
  {de~Lasson}, \citenamefont {Kristensen}, \citenamefont {M{\o}rk},\ and\
  \citenamefont {Gregersen}}]{deLasson}%
  \BibitemOpen
  \bibfield  {author} {\bibinfo {author} {\bibfnamefont {J.~R.}\ \bibnamefont
  {de~Lasson}}, \bibinfo {author} {\bibfnamefont {P.~T.}\ \bibnamefont
  {Kristensen}}, \bibinfo {author} {\bibfnamefont {J.}~\bibnamefont {M{\o}rk}},
  \ and\ \bibinfo {author} {\bibfnamefont {N.}~\bibnamefont {Gregersen}},\
  }\href {\doibase 10.1364/OL.40.005790} {\bibfield  {journal} {\bibinfo
  {journal} {Opt. Lett.}\ }\textbf {\bibinfo {volume} {40}},\ \bibinfo {pages}
  {5790} (\bibinfo {year} {2015})}\BibitemShut {NoStop}%
\bibitem [{Note1()}]{Note1}%
  \BibitemOpen
  \bibinfo {note} {The calculated $S$ factors are: $S_{11}{=}0.7485 + 0.7356$,
  $S_{22}{=}0.8006+0.4409$, $S_{12}{ = }({-}0.0043 {-} 0.5574i)+(0.0579{-}
  0.4137i)$, and $S_{21}{=}S^*_{12}$, where first and second parts denote
  $S_{\mu \eta }^{\protect \rm nrad}$ and $S_{\mu \eta }^{\protect \rm rad}$,
  respectively.}\BibitemShut {Stop}%
\end{thebibliography}
\end{document}